\documentclass[conference]{IEEEtran}

\usepackage{xcolor}
\usepackage{xspace}
\usepackage{subcaption}
\usepackage{hyperref}
\usepackage{multirow}
\usepackage{setspace}
\usepackage[ruled,vlined]{algorithm2e}
\usepackage{algpseudocode}
\usepackage{glossaries}
\usepackage{booktabs}
\usepackage{graphicx}
\usepackage{tabularray}
\usepackage{blkarray}
\usepackage{fontawesome5}
\usepackage{todonotes}
\usepackage{circledsteps}
\usepackage{cite}
\usepackage{url}
\usepackage{listings}
\usepackage{amsfonts}
\hypersetup{
    pdfborder={0 0 0}
}

\newcommand{\ie}{i.e.,\xspace}
\newcommand{\eg}{e.g.,\xspace}

\newcommand{\etal}{{et al.}\xspace}

\newcommand{\apk}{{APK}\xspace}
\newcommand{\apks}{{APKs}\xspace}
\newcommand{\practicaldroid}{\textsc{DroidBreaker}\xspace}
\newcommand{\advdroidzero}{\textsc{AdvDroidZero}\xspace}
\newcommand{\ade}{\textsc{ADE}\xspace}
\newcommand{\drebin}{{Drebin}\xspace}
\newcommand{\mamadroid}{{MaMaDroid}\xspace}
\newcommand{\apigraph}{{APIGraph}\xspace}
\newcommand{\secsvm}{{SecSVM}\xspace}
\newcommand{\elsa}{{ELSA}\xspace}
\newcommand{\apg}{{APG}\xspace}

\newcommand{\funrate}{Functional Rate\xspace}
\newcommand{\drebinred}{{Drebin-10k}\xspace}
\newcommand{\secsvmred}{{SecSVM-10k}\xspace}

\usepackage{lipsum}
\makeatletter
\newcommand{\algorithmfootnote}[2][\footnotesize]{%
  \let\old@algocf@finish\@algocf@finish% Store algorithm finish macro
  \def\@algocf@finish{\old@algocf@finish% Update finish macro to insert "footnote"
    \leavevmode\rlap{\begin{minipage}{\linewidth}
    #1#2
    \end{minipage}}%
  }%
}

\newcommand{\vct}[1]{\ensuremath{\boldsymbol{#1}}}
\newcommand{\mat}[1]{\ensuremath{\mathbf{#1}}}
\newcommand{\set}[1]{\ensuremath{\mathcal{#1}}}

\newcommand{\T}{\ensuremath{\top}}
\newcommand{\myparagraph}[1]{\noindent\textbf{#1.}}

\newcommand{\argmin}{\operatornamewithlimits{\arg\,\min}}

\SetKwInOut{Input}{Input}
\SetKwInOut{Output}{Output}

\newcommand{\yes}{\scalebox{0.7}{\faCheck}\xspace}
\newcommand{\no}{\scalebox{0.7}{\faTimes}\xspace}
\newcommand{\rem}{\scalebox{0.7}{\faCircle}\xspace}
\newcommand{\inj}{\scalebox{0.7}{\faCircle[regular]}\xspace}
\newcommand{\injrem}{\scalebox{0.7}{\faAdjust}\xspace}

\newcommand{\low}{{\tt LOW}\xspace}
\newcommand{\med}{{\tt MED}\xspace}
\newcommand{\high}{{\tt HIGH}\xspace}
\newcommand{\man}{\scalebox{0.7}{\faHandPaper}\xspace}

\newcommand{\bb}{\scalebox{0.7}{\faSquare}\xspace}
\newcommand{\wb}{\scalebox{0.7}{\faSquare[regular]}\xspace}

\DeclareFontEncoding{LS1}{}{}
\DeclareFontSubstitution{LS1}{stix}{m}{n}
\DeclareSymbolFont{stixsym}{LS1}{stixscr}{m}{n}
\DeclareMathSymbol{\squarelrblack}{\mathord}{stixsym}{"FF}
\newcommand{\bbwb}{\raisebox{1.0pt}{\scalebox{0.8}{$\squarelrblack$}}\xspace}

\newcommand{\blackcircle}[1]{{\small \Circled[fill color=black, inner color=white]{#1}}}

\newcommand{\comp}[1]{$\set C_{#1}$}
\definecolor{mygreen}{rgb}{0,0.6,0}
\definecolor{mygray}{rgb}{0.5,0.5,0.5}
\definecolor{mymauve}{rgb}{0.58,0,0.82}

\newcommand{\codehl}[2]{%
{\setlength{\fboxsep}{1pt}\colorbox{#1}{#2}}%
}
\lstdefinestyle{smali}{
  language=[x86masm]Assembler, % Use Assembly as a base
  basicstyle=\ttfamily\scriptsize,
  keywordstyle=\color{black}\bfseries,
  stringstyle=\color{black},
  commentstyle=\color{black},
  backgroundcolor=\color{black!5},
  showstringspaces=false,
  frame=single,
  breaklines=true,
  tabsize=4,
  morekeywords={}
}
\lstdefinestyle{xml}{
  language=xml,
  basicstyle=\ttfamily\scriptsize,
  keywordstyle=\color{black}\bfseries,
  stringstyle=\color{black},
  commentstyle=\color{black},
  backgroundcolor=\color{black!5},
  showstringspaces=false,
  frame=single,
  breaklines=true,
  tabsize=4,
  morekeywords={}
}

\begin{document}
\newacronym{CFG}{CFG}{Control Flow Graph}
\newacronym{FCG}{FCG}{Function Call Graph}
\newacronym{VT}{VT}{VirusTotal}
\newacronym{AV}{AV}{AntiVirus}
\newacronym{UI}{UI}{User Interface}

\title{DroidBreaker: Practical and Functional Problem-Space Attacks on Machine-Learning Android Malware Detectors}

% PAGE LIMITS: 13 (excluding Ethical) +5

%\author{Anonymous Authors}

\author{
\IEEEauthorblockN{
   Christian Scano\IEEEauthorrefmark{1}\IEEEauthorrefmark{2},
   Diego Soi\IEEEauthorrefmark{1},
   Angelo Sotgiu\IEEEauthorrefmark{1}\IEEEauthorrefmark{3},
   Luca Demetrio\IEEEauthorrefmark{4},
   Davide Maiorca\IEEEauthorrefmark{1},\\
   Giorgio Giacinto\IEEEauthorrefmark{1}\IEEEauthorrefmark{3},
   Fabio Roli\IEEEauthorrefmark{1}\IEEEauthorrefmark{3}\IEEEauthorrefmark{4}, and
   Battista Biggio\IEEEauthorrefmark{1}\IEEEauthorrefmark{3},
}
\IEEEauthorblockA{\IEEEauthorrefmark{1}University of Cagliari}
\IEEEauthorblockA{\IEEEauthorrefmark{2}Sapienza University of Rome}
\IEEEauthorblockA{\IEEEauthorrefmark{3}CINI}
\IEEEauthorblockA{\IEEEauthorrefmark{4}University of Genova}
\{christian.scano, diego.soi, angelo.sotgiu, davide.maiorca, giorgio.giacinto, battista.biggio\}@unica.it,\\
\{luca.demetrio, fabio.roli\}@unige.it
}
	
% \IEEEoverridecommandlockouts
% \makeatletter\def\@IEEEpubidpullup{6.5\baselineskip}\makeatother
% \IEEEpubid{\parbox{\columnwidth}{
% 		Network and Distributed System Security (NDSS) Symposium 2026\\
% 		23 - 27 February 2026 , San Diego, CA, USA\\
% 		ISBN 979-8-9919276-8-0\\  
% 		https://dx.doi.org/10.14722/ndss.2026.[23$|$24]xxxx\\
% 		www.ndss-symposium.org
% }
% \hspace{\columnsep}\makebox[\columnwidth]{}}

\maketitle

\begin{abstract}
Adversarial APKs are Android applications modified in the problem space to evade machine‑learning malware detectors. In this work, we first show that, despite claims, existing problem-space attacks remain largely impractical. Most techniques leverage software transplantation to inject entire benign modules, introducing many side-effect features and often causing build-time failures. Fine-grained methods that inject only a narrow subset of components exhibit limited effectiveness, while those that also use obfuscation rely on brittle bytecode rewriting, producing APKs that are syntactically valid but semantically unusable. Prior work further overestimates attack success rates by running smoke tests that only validate installation and basic execution, without assessing whether the modified APK still preserves its intended behavior.
To overcome these limitations, we present \practicaldroid, a practical (build-safe) and functional (semantics‑preserving) problem‑space attack framework that provides: (i)~query-efficient white- and black-box attacks by manipulating only the APK components most influential to the target model; (ii)~a set of fine-grained, build-safe manipulations (including injection and obfuscation of API calls, app modules, permissions, and URLs) with minimal side effects; and (iii) a semantics-preserving functionality test that enforces runtime equivalence by comparing execution logs and API-level traces between the initial and the modified APK.
Evaluated on a recent corpus of Android applications, \practicaldroid achieves high evasion rates with few queries and minimal side effects in both white‑box and black‑box settings, and drastically reduces detections by commercial malware scanners hosted on VirusTotal.
\end{abstract}


% \fi
\IEEEpeerreviewmaketitle

\section{Introduction}
Android is the most widely deployed mobile operating system, with billions of devices and millions of applications across both official and third-party app stores.
This scale makes it a prime target for malicious \apks that infiltrate devices to steal data, credentials, or financial information.
Malware detectors based on machine learning (ML) have been widely adopted to counter these threats, leveraging large training datasets collected from diverse sources~\cite{ruggia2024unmasking,arp2014drebin,allix2016androzoo} and exploring a variety of feature representations and models~\cite{arp2014drebin,demontis2017yes,mariconti2017mamadroid}.
However, their effectiveness remains limited. Their performance decays quickly as malware evolves over time~\cite{pendlebury2019tesseract}, and they remain vulnerable to adversarial manipulation of the \apk files~\cite{grosse2017adversarial,demontis2017yes,pierazzi2020intriguing,li2020adversarial,yang2017malware,xu2023gendroid,he2023efficient,Bostani2024_COSE}. %qui ne mancano
Early work demonstrated this vulnerability via \textit{feature-space} attacks, \ie by only changing a small subset of feature values among those provided as input to the classifier, without crafting the corresponding adversarial APKs~\cite{grosse2017adversarial,demontis2017yes}. Subsequent research has thus questioned their practical applicability and shifted the focus towards \textit{problem-space} attacks, \ie attacks that produce functional adversarial APKs by manipulating components such as bytecode, resources, or manifest entries~\cite{pierazzi2020intriguing,li2020adversarial,yang2017malware,Bostani2024_COSE}.

In this work, we show that existing problem-space attacks for Android malware detection are far less practical than claimed: \textit{coarse-grained} injection methods that rely on \emph{software transplantation}~\cite{barr2015automated}  inject entire portions of benign APKs into malicious ones, introducing many unintended side-effect features and often failing to repackage, causing build-time failures~\cite{yang2017malware, pierazzi2020intriguing, he2023efficient, Bostani2024_COSE};
\textit{fine-grained} methods that inject only a smaller subset of components, such as APIs and Permissions, exhibit limited effectiveness~\cite{shu2024eagle,xu2023gendroid,chen2019android,Li2023_Usenix}, 
while those that also perform code obfuscation rely on brittle bytecode rewriting, frequently breaking app semantics~\cite{li2020adversarial}. 
Furthermore, we highlight another significant prior work limitation, related to how they test that attacks preserve the initial \apk malicious behavior. In particular, prior evaluations validate that the app semantics are preserved using smoke tests that only check installation and basic execution~\cite{xu2023gendroid, shu2024eagle,pierazzi2020intriguing, chen2019android, yang2017malware}, 
without really assessing whether the modified APK preserves its intended functionality, e.g., by looking at its execution traces when properly stimulated at runtime. We will show that this leads to significantly overestimating attack success rates. 
For these reasons, we argue that the problem of crafting \textit{practical} (build-safe) and \textit{functional} (semantics-preserving) problem‑space attacks still represents an open challenge (\autoref{sec:background}).
By \textit{practical}, we refer to attacks that can be reliably applied without failing to repackage, while by \textit{functional}, we refer to attacks that avoid runtime crashes and preserve the app semantics.
Finally, existing approaches also lack \textit{scalability}: unstable manipulations,  manual validation, and insufficient functionality testing prevent the reliable generation of attacks and the preservation of behavior at scale.

To overcome these issues, we present \practicaldroid (\autoref{sec:methodology}), a framework for designing \textit{practical} and \textit{functional} problem-space attacks, providing the following contributions.

\myparagraph{(1) Query-efficient white-box and black-box attacks} \practicaldroid can be used to stage both white- and black-box attacks that can drastically reduce the number of queries to the target model (i.e., forward and backward passes). To this end, we define an attack initialization phase whose goal is to retain only manipulations that (i) do not cause build-time failures, and (ii) cause a significant impact on the target model.
We also provide an additional contribution, introducing an \textit{encoding trick} to show that problem-space attacks can be optimized end-to-end via gradient descent, while elegantly formalizing the presence of side effects~\cite{pierazzi2020intriguing}. 

\myparagraph{(2) Practical APK manipulations, minimal side effects} Another significant advancement with respect to existing problem-space attacks is that \practicaldroid performs reliable fine-grained injection and obfuscation of API calls, app modules, Permissions, and URLs using bytecode-rewriting methods and manipulations that are fully compatible with the Android framework, avoid build-time repackaging failures, and introduce only minimal side effects.

\myparagraph{(3) Semantics-preserving functionality testing} To preserve semantics, we introduce new functionality tests enforcing runtime equivalence between initial and modified APKs by comparing execution logs and API‑level traces, validating that our attacks are functional and do not cause runtime errors.

Our experiments (\autoref{sec:evaluation}) demonstrate the strong evasive capabilities of \practicaldroid against four detectors with distinct feature representations, consistently achieving high evasion rates within only a few queries and with minimal side effects. We further show that the adversarial \apks generated by \practicaldroid also impact commercial scanners, reducing detections by roughly half on \gls{VT}~\cite{virustotal}.
This shows that, for the first time, practical and functional problem-space attacks against Android malware detectors are not only possible but effective at scale. 
We conclude the paper by summarizing our work, discussing its limitations, and providing future research directions in~\autoref{sec:conclusions}. 
% To foster reproducibility, we open-source \practicaldroid implementation at \url{https://anonymous.4open.science/r/droidbreaker}.

\section{Adversarial APKs: Are We There Yet?}
\label{sec:background}

\begin{table*}[t]
\small
\caption{Problem-space attacks for adversarial APKs:  \blackcircle{1} Attack Setting (\xspace\wb white-box, \bb black-box, or \bbwb both); \blackcircle{2} APK Manipulations (\inj injection, \rem obfuscation, or \injrem both) and affected components (\comp{1}-\comp{6}), with qualitative evaluation of  \textit{practical} (build-safe) manipulations (\yes yes, \no no, - not verifiable), \textit{side effects} and expected \textit{effectiveness} on different detectors; \blackcircle{3} Functionality Testing (\man manual, \yes automatic, - not applied), with details of \textit{smoke} and \textit{runtime} testing phases, and rate of \textit{functional} (semantics-preserving) apps; and \blackcircle{4} Open Code availability of attack and functionality tests (\yes yes, \no no).}
\label{tab:att_comparison}
\centering
\resizebox{\textwidth}{!}{
\begin{tblr}{
    colspec = {l l c c ccccccccc cccccc cc},
    rows = {rowsep = 2pt},
    row{15} = {bg = black!7},
    % --- LINEE ORIZZONTALI GENERALI ---
    hline{1,Z} = {solid, wd=1pt}, 
    % --- LINEE PARZIALI ---
    hline{2} = {5-13}{solid, leftpos = -0.5, rightpos = -0.5, endpos},
    hline{2} = {14-19}{solid, leftpos = -0.5, rightpos = -0.5, endpos},
    hline{2} = {20-21}{solid, leftpos = -0.5, rightpos = -0.5, endpos},
    hline{3} = {14-15}{solid, leftpos = -0.5, rightpos = -0.5, endpos},
    hline{3} = {16-18}{solid, leftpos = -0.5, rightpos = -0.5, endpos},
    % --- LINEE VERTICALI ---
    vline{3,4,5} = {solid},
    vline{11,12,13,14,16,19,20} = {solid},
}
    % --- RIGA 1 ---
    & & & & \SetCell[c=9]{c} \textbf{{\small \blackcircle{2}} APK Manipulations} & & & & & & & & & 
      \SetCell[c=6]{c} \textbf{{\small \blackcircle{3}} Functionality Testing} & & & & & & 
      \SetCell[c=2]{c} \textbf{{\small \blackcircle{4}} Open Code} & \\

    % --- RIGA 2 ---
    & & & & & & & & & & & & &
      \SetCell[c=2]{c} Smoke & & 
      \SetCell[c=3]{c} Runtime & & & & & \\
    
    % --- RIGA 3: Headers ---
    & \textbf{Attack} 
      & \rotatebox{90}{\textbf{Year (Preprint)}} 
      & \rotatebox{90}{\textbf{{\small \blackcircle{1}} Attack Setting}} 
      & \rotatebox{90}{$C_1$: App Modules}
      & \rotatebox{90}{$C_2$: Hw. Features}
      & \rotatebox{90}{$C_3$: Permissions}
      & \rotatebox{90}{$C_4$: Intent Filters}
      & \rotatebox{90}{$C_5$: APIs}
      & \rotatebox{90}{$C_6$: Strings}
      & \rotatebox{90}{\textbf{Practical}}
      & \rotatebox{90}{\textbf{Side Effects}}
      & \rotatebox{90}{\textbf{Effectiveness}}
      & \rotatebox{90}{Repackaging}
      & \rotatebox{90}{Install}
      & \rotatebox{90}{Launch}
      & \rotatebox{90}{Exec. Logs}
      & \rotatebox{90}{Exec. Traces}
      & \rotatebox{90}{\textbf{Functional}}
      & \rotatebox{90}{\textbf{Attack}}
      & \rotatebox{90}{\textbf{Functionality Test}} \\
    
    \hline \hline
    % --- Gruppo 1: soft. transpl. ---
    \SetCell[r=4]{m}{\rotatebox{90}{\small\textit{Coarse}}} 
      & Yang~\etal~\cite{yang2017malware} 
      & 2017 & \bb 
      & \inj & \inj & \inj & \inj & \inj & \inj 
      & \no & \high & \med
      & \yes & \yes & - & - & -
      & \low
      & \no & \no \\
    & Pierazzi~\etal~\cite{pierazzi2020intriguing}
      & 2020 & \wb
      & \inj & \inj & \inj & \inj & \inj & \inj 
      & \no & \high & \med
      & \yes & \man & \man & - & -
      & \low
      & \yes & \no \\
    & \textsc{EvadeDroid}~\cite{Bostani2024_COSE} 
      & 2021 & \bb
      & \inj & \inj & \inj & \inj & \inj & \inj 
      & \no & \high & \med
      & \yes & \yes & \yes & \yes & - 
      & \med
      & \yes & \no \\
    & \advdroidzero~\cite{he2023efficient} 
      & 2023 & \bb
      & \inj & \inj & \inj & \inj & \inj & \inj 
      & \no & \med & \med
      & \yes & \man & \man & \man & -
      & \med
      & \yes & \no \\

    \hline \hline
    % --- Gruppo 2: fine ---
    \SetCell[r=6]{m}{\rotatebox{90}{\small\textit{Fine}}} 
      & \textsc{Android HIV}~\cite{chen2019android} 
      & 2018 & \wb
      & - & - & \inj & - & \inj & \inj
      & - & \low & \low
      & \yes & \man & \man & - & -
      & \high
      & \no & \no \\
    & \textsc{HRAT}~\cite{zhao2021structural} 
      & 2021 & \wb
      & - & - & - & - & \injrem & -
      & \no & \low & \low
      & \yes & \man & \man & \man & - 
      & \high
      & \yes & \no \\
    & \textsc{BagAmmo}~\cite{Li2023_Usenix} 
      & 2023 & \bb
      & - & - & - & - & \inj & -
      & - & \low & \low
      & \yes & \man & \man & \man & -
      & \high
      & \no & \no \\ 
    & \textsc{GenDroid}~\cite{xu2023gendroid} 
      & 2023 & \bb
      & - & - & - & - & \inj & -
      & - & \low & \low
      & \yes & - & - & - & -
      & \high
      & \no & \no \\
    & \textsc{EAGLE}~\cite{shu2024eagle} 
      & 2023 & \bb 
      & - & - & \inj & - & \injrem & -
      & - & \low & \low
      & \yes & - & - & - & -
      & \high
      & \no & \no \\ 
    
    \hline

    & \ade~\cite{li2020adversarial} 
      & 2020 & \wb
      & \injrem & \inj & \inj & \inj & \injrem & \injrem
      & \no & \low & \high
      & \yes & \yes & \yes & - & -
      & \low
      & \yes & \no \\ 
    
    \hline\hline
    
    & \practicaldroid (ours)
      & - & \bbwb
      & \injrem & \inj & \inj & \inj & \injrem & \injrem
      & \yes & \low & \high
      & \yes & \yes & \yes & \yes & \yes
      & \high
      & \yes & \yes \\
\end{tblr}}
\end{table*}

Previous work on problem-space attacks may give the impression that crafting \textit{functional} adversarial \apks, \ie malicious applications that evade detection while preserving runtime behavior, is a solved problem~\cite{yang2017malware,pierazzi2020intriguing,he2023efficient,Bostani2024_COSE,shu2024eagle,xu2023gendroid,chen2019android,Li2023_Usenix,zhao2021structural,li2020adversarial}.
In this work, we demonstrate that ensuring semantic preservation under adversarial manipulation remains a challenging issue, far from being solved. To this end, we first categorize existing attacks in~\autoref{tab:att_comparison} along four main dimensions, detailed below. 

\myparagraph{\blackcircle{1} Attack Setting} It captures whether the attack is staged in a white- (\wb) or black-box (\bb) setting, or in both (\bbwb) settings.

\myparagraph{\blackcircle{2} APK Manipulations} This dimension reports the components \comp{1}-\comp{6} (cf. \autoref{subsec:adv_apk:manipulation}) modified by the attack via injection (\inj), obfuscation (\rem), or both (\injrem). We also report: (i) whether the attack is \textit{practical}, i.e., build safe (\yes), or not (\no), along with a qualitative assessment of (ii) the amount \textit{side-effect features} (\high for coarse-grained injection methods, and \low for fine-grained ones); and (iii) the expected \textit{effectiveness} against diverse detectors (\low for attacks manipulating few components, \med for coarse-grained injection attacks manipulating many components, and \high for attacks combining injection and obfuscation over a large set of components). 

\myparagraph{\blackcircle{3} Functionality Testing} It characterizes the depth of APK functionality validation after manipulation (cf. \autoref{subsec:adv_apk:funct_test}), ranging from \textit{smoke} to \textit{runtime} testing, and indicates whether validation is automated (\yes) or manual (\man).
We also qualitatively report the \textit{functional} rate after attack (\low, \med, \high), defined as the fraction of adversarial APKs that preserve the intended behavior. 
For the attacks considered in our experiments, we report the exact values in \autoref{sec:evaluation}. 

\myparagraph{\blackcircle{4} Open Code} It highlights whether \textit{attack} and corresponding \textit{functionality test} code is publicly available (\yes) or not (\no), impacting its reproducibility and re-evaluation.

Based on this analysis, we identify fundamental issues undermining current attacks.
In particular, many of them fail to preserve the intended behavior and consequently overestimate attack success rates, as they rely on inconsistent or faulty manipulations (\autoref{subsec:adv_apk:manipulation}) and incomplete functionality tests (\autoref{subsec:adv_apk:funct_test}). We discuss these limitations in detail in \autoref{subsec:adv_apk:sum_limitation}, highlighting that only a few attack implementations are open-source, thereby hindering reproducibility and re-evaluation.

\subsection{APK Manipulations}
\label{subsec:adv_apk:manipulation}
We categorize here prior APK manipulations based on the affected components and the strategies used to modify the APK. Further details are provided in Appendix~\ref{appendix:problem-space} and~\ref{app:manipulation:types}.

\subsubsection{\apk Components} 
\label{subsubsec:adv_apk:apk_components}
APKs are compressed archives consisting of: (i)~the \emph{manifest}, declaring the app structure and its required capabilities to the Android OS; (ii)~the \emph{code} (DEX bytecode, disassembled into smali code, and native libraries), implementing the functionalities referenced in the manifest; and (iii)~the \emph{resources}, including non-code app elements, such as \gls{UI} layouts, images, strings, and styles.
We report below six categories of APK components (\comp{1}-\comp{6}) that are manipulated by current attacks, as detailed in \autoref{tab:att_comparison}. Examples for each category are reported in Appendix~\ref{app:app_components}.

\myparagraph{\comp{1}: Application Modules} They correspond to the class names of Activities, Services, Receivers, and Providers declared in the manifest and implemented inside the \apk DEX bytecode.

\myparagraph{\comp{2}: Hardware Features} They are declared in the manifest and grant permission to use specific hardware (\eg mic, camera).

\myparagraph{\comp{3}: Permissions} They are declared in the manifest and grant access to specific software capabilities (\eg sending SMS).
   
\myparagraph{\comp{4}: Intent Filters} They are reported in the manifest and describe the external behaviors of invoked app modules.

\myparagraph{\comp{5}: APIs} They allow to interact with the Android~OS.

\myparagraph{\comp{6}: Strings} Strings (\eg URLs, IPs) defined in the bytecode.

\subsubsection{Manipulation Strategies}
\label{subsubsec:adv_apk:manipulation}
We cover the main methods to craft adversarial APKs: coarse-grained injection via software transplantation and fine-grained injection and obfuscation.

\myparagraph{Coarse-grained Injection (Software Transplantation)} The first set of attacks in \autoref{tab:att_comparison}~\cite{yang2017malware, pierazzi2020intriguing, he2023efficient, Bostani2024_COSE} relies on \emph{software transplantation}~\cite{barr2015automated}, a technique originally proposed to automatically migrate functionalities between software systems. It operates by identifying a target component from a donor, harvesting all its necessary dependencies, and injecting the resulting code slice into the host.
In the context of adversarial \apks, software transplantation was first introduced in~\cite{yang2017malware} to migrate components (\eg API calls and App Modules) across malicious APKs, and extended in subsequent work~\cite{pierazzi2020intriguing,Bostani2024_COSE,he2023efficient} to inject components from benign \apks into malware samples. More specifically, Pierazzi~\etal~\cite{pierazzi2020intriguing} targets Activities, Receivers, Providers, URLs, API calls, and Permissions; \textsc{EvadeDroid}~\cite{Bostani2024_COSE} only focuses on API calls; and \advdroidzero~\cite{he2023efficient} targets Services, Receivers, and Providers (while also injecting Hardware Features, Permissions, and Intent Filters in the manifest file).
By construction, the process required to transplant Android components results in the implantation of large APK portions (typically the entire implementation of app modules and their associated dependencies and manifest entries), thus enabling the injection of all APK components listed in \autoref{tab:att_comparison} (\comp{1}-\comp{6}).
These components are added without altering the execution path by injecting code that is not executed at runtime, claiming to preserve the \apk semantics by design~\cite{pierazzi2020intriguing}.
However, while effective in principle, this approach introduces a large number of unintended side-effect components that uncontrollably increase the complexity of the resulting \apks. As a consequence, the manipulation process becomes highly unstable and unreliable, especially when many components need to be injected, often leading to malformed or non-functional applications. 
Our experimental evaluation in \autoref{sec:evaluation} shows that these approaches are largely impractical, and achieve only a \textit{medium} expected effectiveness against diverse detectors.
The reason is that they generate significantly suboptimal adversarial \apks, as these manipulations introduce \textit{too many} uncontrolled side‑effect features. This in turn leads to overly optimistic evaluations, which only contribute to spreading a false sense of security~\cite{carlini2019evaluating}.

\myparagraph{Fine-grained Injection and Obfuscation} The second set of attacks in \autoref{tab:att_comparison}~\cite{shu2024eagle, xu2023gendroid, chen2019android, Li2023_Usenix, zhao2021structural} apply fine‑grained manipulations on a restricted subset of APK components, \ie Permissions, APIs, and Strings.
Limiting modifications to a narrow set of components enables them to introduce fewer side-effect features, resulting in more controlled manipulations and a generally higher functional rate than coarse-grained injection attacks.
However, their expected effectiveness remains \textit{low} against detectors based on diverse or more comprehensive feature sets, limiting their practical impact.
Among the methods with available code, we verified that \textsc{HRAT}~\cite{zhao2021structural} is not practical. It injects and obfuscates methods that are neither framework APIs nor lifecycle methods, and applies code transformations that may break the \apk build, e.g., moving a method body into its caller. If register handling is incorrect or exceeds the declared limits, the resulting bytecode fails to compile.
\ade~\cite{li2020adversarial} is the only attack in this category that allows injecting and obfuscating multiple components in a fine-grained manner. However, its flawed implementation frequently breaks \apks: its manipulations may (i) leave unresolved references to obfuscated components, causing crashes when they are accessed; (ii) prevent retrieving or invoking the obfuscated elements; and (iii) inject transformed components whose integration accidentally results in runtime failures. 

\myparagraph{\underline{A Paradigmatic Example}} While we defer a detailed technical analysis of all these issues to~\autoref{app:manipulation:ade}, we report in \autoref{fig:eg-fault} an example of a URL (string) injection via software transplantation~\cite{pierazzi2020intriguing} (\autoref{fig:injection-string-pierz}), and via \ade~\cite{li2020adversarial} (\autoref{fig:injection-string-ade}).
In the first case, the entire class containing the target URL is added, introducing additional dependencies in the form of stub classes (\eg \texttt{android.telephony.TelephonyManager}) as side effects to satisfy type references (via FlowDroid~\cite{arzt2014flowdroid}) and unnecessary components for the attack (\eg \texttt{setRequestMethod}).  
In the second case, the URL is injected within a register that cannot hold strings, resulting in invalid bytecode that breaks the APK during either building or verification. Our approach, instead, overcomes these limitations by minimizing uncontrolled side effects and providing practical (build-safe) and functional (semantics-preserving) manipulations (cf. \autoref{fig:injection-our}, \autoref{subsec:methodology:func_pres_manipulation}).

\begin{figure}[t]
\centering
    \begin{subfigure}{0.46\textwidth}
        \centering
        \begin{lstlisting}[style=smali,escapechar=?]
.class public final Lcom/inmobi/ads/c;
.super Lcom/inmobi/commons/core/configs/a;
.method static constructor <clinit>()V
    .registers 8
    [...]
    const-string v4, "?\codehl{blue!20}{https://sdktm.w.inmobi.com}?"
    [...]
    const-string v5, "GET"
    new-instance v0, Ljava/net/HttpURLConnection;
    invoke-virtual {v0, v5}, ?\codehl{red!20}{Ljava/net/HttpURLConnection;->}?
    ?\codehl{red!20}{setRequestMethod(Ljava/lang/String;)V}?
    [...]
.end method

.class public Landroid/telephony/TelephonyManager;
.super Ljava/lang/Object;
.method public getDeviceId()Ljava/lang/String;
    .registers 3
    new-instance v0, Ljava/lang/RuntimeException;
    .local v0, "$r1":Ljava/lang/RuntimeException;, ""
    const-string v1, "Stub!"
    invoke-direct {v0, v1}, Ljava/lang/RuntimeException;-><init>(Ljava/lang/String;)V
    throw v0
    .end local v0    
.end method
\end{lstlisting}
        \vspace{-2\baselineskip}
        \caption{String injection via software transplantation~\cite{pierazzi2020intriguing}.}
        \label{fig:injection-string-pierz}
    \end{subfigure}
    \hfill
    \begin{subfigure}{0.46\textwidth}
        \begin{lstlisting}[style=smali,escapechar=?]
.method public static <init>()V
    .registers 1
    const-string p0, "?\codehl{blue!20}{https://sdktm.w.inmobi.com}?" ?\label{line:url_inj}?
    iput-object p1,p0,Ldj;->a:Lcom/m_zxmlmnnew/image/pc8aa4a7efb;
    invoke-direct {p0}, Ljava/lang/Object;-><init>()V
    return-void
.end method\end{lstlisting} 
    \vspace{-2\baselineskip}
    \caption{String injection via \ade~\cite{li2020adversarial}.}
    \label{fig:injection-string-ade}
    \vspace{-5pt}
    \end{subfigure}
    \caption{Examples of URL (highlighted in blue) string injection. In (a), the entire class {\tt TelephonyManager} and an API call (in red) are injected as a side effect; in (b), string injection breaks the \apk build due to faulty register management.}
    \label{fig:eg-fault}
\end{figure}

\subsection{Functionality Testing}
\label{subsec:adv_apk:funct_test}
Adversarial \apks must fully preserve their intended malicious behavior after manipulation. We review here how prior work conducted \emph{functionality testing} to validate that. As summarized in~\autoref{tab:att_comparison}, existing approaches can be broadly categorized into two families, \ie \textit{smoke} and \textit{runtime} testing.

\myparagraph{Smoke Testing} This preliminary validation step checks whether basic structural integrity has been preserved. Typically, it verifies that adversarial \apks can be successfully repackaged
%(\autoref{app:problem-space:repackaging})
and installed on emulators or real devices.
As shown in~\autoref{tab:att_comparison}, this is often applied in a minimal and inconsistent manner throughout prior work. Some attacks only consider repackaging~\cite{xu2023gendroid,shu2024eagle}, while most also verify that adversarial \apks can be installed without errors. However, installation tests are typically performed on a small number of samples and frequently rely on manual inspection, providing only weak evidence of the correctness and robustness of the manipulation process.
While install failures may indicate severe corruption of the \apk, smoke testing does not assess semantics preservation, as the app is never executed. Thus, it ignores runtime errors, crashes, and behavioral deviations that may only manifest during execution, providing no evidence that the intended functionality is actually preserved.

\myparagraph{Runtime Testing} It aims to assess functionality preservation by observing the execution of adversarial \apks in sandboxes~\cite{CuckooDroid} or emulators~\cite{li2020adversarial, Li2023_Usenix}. In prior work, this is typically achieved by verifying that \apks can be launched and executed without crashing (\emph{Launch} in~\autoref{tab:att_comparison}), and by also checking the presence of specific implanted messages in the \texttt{Logcat} output (\emph{Execution Logs} in~\autoref{tab:att_comparison}).\footnote{\texttt{Logcat} is a utility recording runtime debugging and error messages.}
In both cases, execution may be stimulated using manual \gls{UI} inspection~\cite{pierazzi2020intriguing, he2023efficient, chen2019android, zhao2021structural,Li2023_Usenix}, or automatic monkey testing~\cite{zhao2021structural, Bostani2024_COSE} simulating random interactions with the app \gls{UI}.
Despite being more informative than smoke testing, these checks still provide only limited guarantees of preserved semantics. While they can confirm that \apks can be executed,\footnote{The Android OS enforces compliance with the Android Verifier~\cite{AndroidVerifier} which checks low‑level bytecode correctness at class loading time.} or that specific log messages are produced, they cannot verify that the initial malicious behavior is fully preserved. This limitation is critical given that malicious logic often executes as background services, without triggering \gls{UI} interactions or producing observable log events. Moreover, most existing approaches rely on manual inspection, small‑scale execution tests, or purely random \apk interaction with the Android monkey tool~\cite{Android_Monkey}, suffering from poor code coverage and limited reproducibility of \gls{UI} events. This severely limits the ability to conduct systematic, reproducible assessments of semantic preservation at scale.

\subsection{What Is Missing?}
\label{subsec:adv_apk:sum_limitation}

From our analysis, we identify two fundamental limitations preventing existing problem-space attacks from being both \emph{practical} (build-safe) and \emph{functional} (semantics-preserving).

\myparagraph{L1: Ineffective/Unreliable Manipulations}
Existing attacks rely on manipulation strategies that are either too coarse or too limited in scope. Coarse-grained approaches based on software transplantation~\cite{yang2017malware, pierazzi2020intriguing, he2023efficient, Bostani2024_COSE} inject entire components and their dependencies, resulting in an uncontrolled inclusion of side-effect features that often destabilize the manipulated APKs, resulting in suboptimal adversarial examples. Fine-grained approaches typically support only injection and/or operate on a small subset of components~\cite{chen2019android,shu2024eagle,xu2023gendroid}, severely limiting their expected effectiveness against diverse detectors. While \ade~\cite{li2020adversarial} also attempts fine-grained obfuscations, its flawed implementation frequently breaks APK repackaging. For these reasons, existing attacks remain largely impractical.

\myparagraph{L2: Inadequate Functionality Testing}
Despite claims of functionality preservation, none of the reviewed works employ functionality tests capable of reliably verifying that adversarial APKs preserve their original behavior. Smoke testing only validates repackaging and installation, while runtime testing typically checks execution without crashes and/or the presence of specific log messages, providing at best weak behavioral evidence for a few manually inspected apps. These checks fail to systematically capture deviations in malicious behavior, which often executes in background services without observable \gls{UI} interactions or logs. As a result, prior work systematically overestimates attack success rates and lacks scalable, sound functionality tests for semantic validation.

\myparagraph{L3: Lack of Open Code} 
Only a limited number of works make their attack implementations publicly available, and even among these, reproducibility is often hampered by bugs and insufficient documentation~\cite{olszewski23getin}. Moreover, \textit{none} of the existing approaches release the code used for functionality testing, hindering reproducibility and independent re-evaluation.

To overcome these limitations, in \autoref{sec:methodology} we introduce \practicaldroid, a novel framework that performs \emph{practical} (build-safe) and \emph{functional} (semantics-preserving) white- and black-box attacks, outperforming competing methods. In particular, \practicaldroid exhibits an improved tradeoff between attack success rate and number of required manipulations, minimizing side-effect features while preserving the intended behavior of adversarial \apks.

\section{\practicaldroid: Practical and Functional Problem-space Attacks for Adversarial APKs}
\label{sec:methodology}

\begin{figure*}[t]
    \centering
    \includegraphics[width=0.9\linewidth]{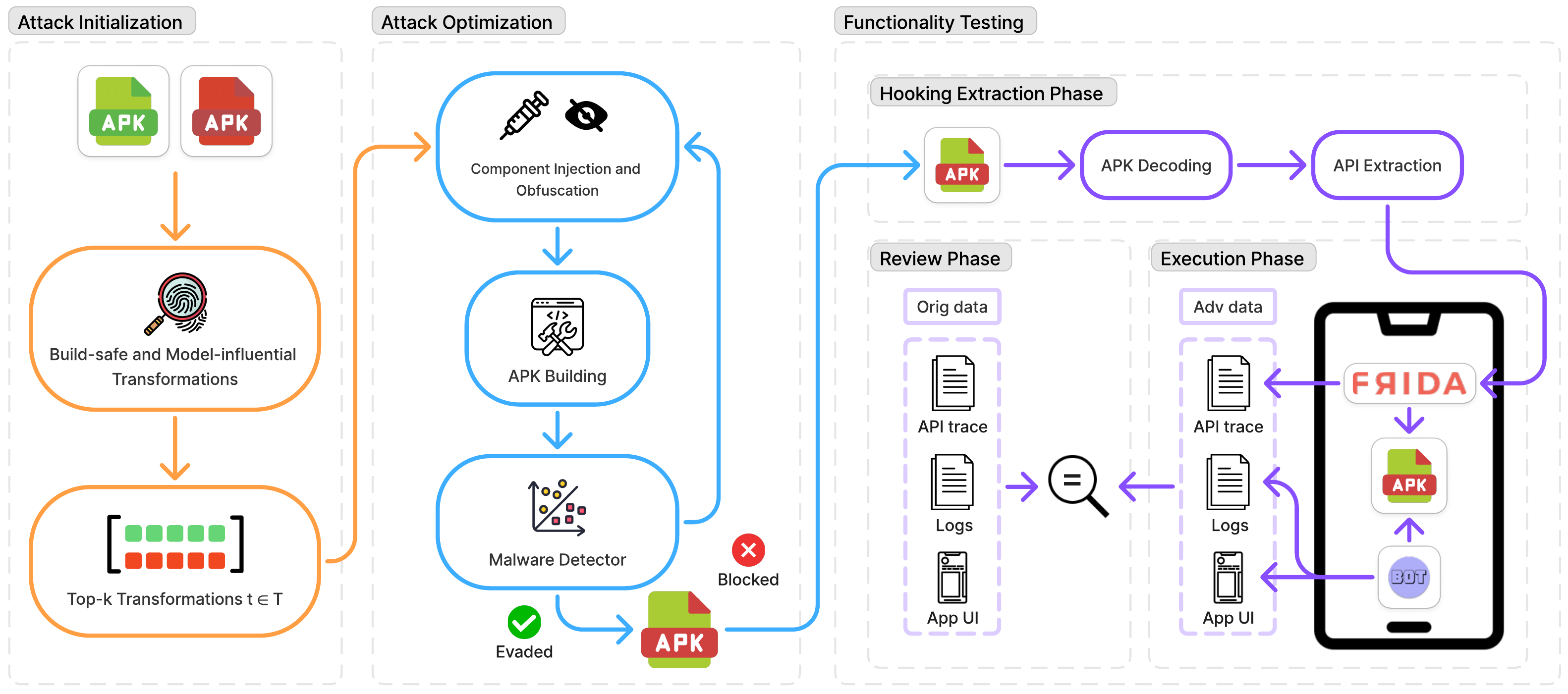}
    \caption{\practicaldroid workflow. (1) Attack Initialization selects the top-k transformations, discarding those that break repackaging or have negligible impact on the model. (2) Attack Optimization manipulates APK components via query-efficient white- and black-box attacks. (3) Functionality Testing uses dynamic analysis to ensure semantics preservation.}
    \label{fig:practicaldroid_arch}
\end{figure*}

We present here \practicaldroid, our novel framework for query-efficient white- and black-box problem-space attacks. Thanks to robust, fine-grained manipulations, it supports both injecting components from benign \apks and obfuscating elements within the target \apk, preserving semantics while minimizing failures and side effects.  
The \practicaldroid workflow, shown in~\autoref{fig:practicaldroid_arch}, consists of three main phases. 
In the \textit{attack initialization} phase (\autoref{subsec:methodology:attack_init}), it selects a reduced set of \apk components eligible for injection or obfuscation, retaining those 
that do not compromise app repackaging and are most relevant to the target
model.
This step is conducted in both white- and black-box attacks to reduce the component space and make optimization significantly more efficient.
The second phase is the \textit{attack optimization} (\autoref{subsec:methodology:attack_opt}), performed either through a gradient-based (white-box) method or a genetic (black-box) algorithm. The third phase is the \textit{functionality testing} (\autoref{subsec:methodology:func_test}), in which \practicaldroid evaluates the functionality of the initial and adversarial \apks using our novel testing procedure to verify that the two samples exhibit the same intended behavior.

\myparagraph{Problem Formulation} Given an input \apk file $z \in \mathcal{Z}$, the detector first extracts a feature vector representation $\vct x = \phi(z) \in\mathcal{X} \subseteq \mathbb{R}^d$. The detector then outputs $f(\vct x) \in \mathbb {R} $, i.e., a confidence value indicating how likely the input sample belongs to the malicious class. The prediction is obtained by thresholding this value: if $f(\vct x) > 0$, the input sample is labeled as malicious, otherwise as legitimate.\footnote{We will use $f(z)$ as shorthand for $f(\phi(z))$ when clear from context.}
In this context, a problem-space attack aims to have an input malware sample misclassified as legitimate by manipulating its components, as defined in \autoref{subsubsec:adv_apk:apk_components}.
The attack can thus be formulated as:
\begin{equation}
\label{eq:attack_generic}
\begin{split}
    \vct t^\star \in \argmin_{\|\vct t\|_0 \leq \lambda} f(\phi(h(z, \vct t))),
\end{split}
\end{equation}
where $h$ is the manipulation function, parameterized by a set of $k$ available transformations that inject/obfuscate components $\vct t \in \set T = \{0,1\}^k$, while $\lambda \in \mathbb{N}$ upper bounds the number of applied transformations.\footnote{While $f$ is defined here as the confidence of the attacked detector, our formulation also holds if one alternatively defines $f$ as a suitable loss function.}
The corresponding adversarial \apk $z^\star$ can then be obtained by applying the optimal transformations $\vct t^\star$ to the initial malware \apk $z$, as $z^\star = h(z, \vct t^\star)$.

\myparagraph{Threat Modeling} We consider attacks that aim to evade detection at test time by manipulating malware samples within the feasible transformation set~$\mathcal{T}$. In the \textit{white-box} case, the attacker knows the feature mapping~$\phi$, the resulting feature vector~$\vct x$, and the scoring function~$f$, enabling direct optimization of Problem~\eqref{eq:attack_generic}. In the \textit{black-box} case, the attacker lacks access to~$\phi$,~$\vct x$, and model parameters, and can only interact with the detector as an oracle by querying modified APKs and observing its output~$f$. In both cases, the attack must satisfy the constraints defined in Section~\ref{subsubsec:adv_apk:apk_components}.

\subsection{Attack Initialization}
\label{subsec:methodology:attack_init}
To be query-efficient, our attacks are initialized by defining the transformation set $\mathcal{T}$ as a subset of $k$ transformations that are both (i)~\textit{build-safe} (preserving correct repackaging) and (ii)~\textit{model-influential} (maximizing impact on the target model). 

\myparagraph{(1) \textit{Build-safe} Transformations} We first construct a set of components by including those that can be injected from a set of benign APKs, and those that can be obfuscated from the malicious input \apk $z$. 
We then exclude from this set the components that, when injected or obfuscated, make the \apk build fail. To this end, we attempt to inject (or obfuscate) all the selected components into the \apk according to the Android structure and context, and compile. 
If compilation/repackaging fails, we split the set into smaller groups and repeat the process until every single problematic component is identified. This step does not require querying the target detector and guarantees that only \textit{build-safe} transformations are kept.

\myparagraph{(2) \textit{Model‑influential} Transformations}\label{subsec:methodology:model_influential} In the second part of the attack initialization phase, we identify which transformations have the largest impact on the detector $f$. In the white-box setting, this requires determining how each transformation influences the features, as described in~\autoref{subsec:methodology:attack_phase_wb}. In the black-box setting, we query the detector by applying all candidate transformations affecting the components in \comp{1}-\comp{6} independently, performing one query per set ($q=6$ in total) to estimate their impact on $f$. We then retain only the top‑ranked transformations, discarding those with no effect. As shown in \autoref{subsubsec:evaluation:model-influential}, this step reduces the transformations from tens to only a few thousand, yielding a compact set $\set T$ of $k$ transformations that improves the attack query efficiency.

\subsubsection*{Practical Manipulations}
\label{subsec:methodology:func_pres_manipulation}
We define here the practical, fine-grained manipulations implemented in \practicaldroid, along with the corresponding modified APK components.
The proposed manipulations are divided into \emph{obfuscation} and \emph{injection}, and applied in a specific order to preserve functionality. Concrete examples are reported in~\autoref{app:manipulation:our}.

\myparagraph{\underline{(1) Component Obfuscation}}
These manipulations are used to obfuscate and increase the source code complexity by modifying class names, strings, and changing the API invocation chain to alter the \gls{FCG}, \ie the graph of the API calls within the program. The primary objective is to hide specific components from the target detector that may fail to recognize their presence within the code.

\myparagraph{App-Module Class Renaming} We replace occurrences of App Modules class identifiers ($\set C_1$) in both the manifest and DEX files with meaningless names. To avoid build-time repackaging failures, we rename only the class names while preserving the original package and class directory structure, without renaming other classes in the same package. This operation is feasible because Android resolves classes through internal identifiers rather than human-readable names, as long as references remain coherent. 

\myparagraph{API Indirection and Reflection} These techniques operate on API calls ($\set C_5$) to modify the \gls{FCG}. First, \textit{indirection} replaces the call to the original method $\gamma_o$ with that to a wrapper method $\gamma_w$. Then, (Java) \textit{reflection} is used to indirectly invoke $\gamma_o$, thereby hiding the API call from the source code.
Unlike the other injection and obfuscation methods, indirection must be applied before reflection to prevent the original method from being hidden by a reflective call, which could cause the application to crash.
Furthermore, to avoid runtime errors, reflection is applied only to declared Android and Java framework APIs, excluding interfaces and abstract methods.

\myparagraph{String Encryption} We encrypt constant strings ($\set C_6$) containing URLs and IP addresses, making it more challenging to identify patterns or the presence of specific strings in the code.

\myparagraph{\underline{(2) Component Injection}}
We support injecting Strings (\ie URLs and IP addresses), API calls, Application Modules, Permissions, and Hardware Features into the original \apks. These components are expected to affect the feature representations of static analysis–based detectors, even if they are not used at runtime. By operating at fine granularity, we inject only a minimal, targeted set of components rather than performing coarse-grained injections, minimizing side‑effect features, build failures, and unintended behavioral changes.

\myparagraph{App-Module Injection} This technique involves inserting additional Activities, Services, Receivers, and Providers ($\set C_1$) with the corresponding Intent Filters ($\set C_4$) in the manifest. To keep the app functional, we decided to set the {\tt enabled} attribute of newly added app modules to {\tt false}, ensuring that the Android verifier at startup does not check for the existence of the corresponding classes inside \apks{}' DEX code.

\myparagraph{Hardware Feature and Permission Injection} We inject Hardware Features ($\set C_2$) and Permissions ($\set C_3$) by adding new entries in the manifest. These injections are inherently semantics-preserving because they do not introduce executable code or runtime behavior that could disrupt functionality. 

\myparagraph{API Injection} We inject API calls ($\set C_5$) into non\-reachable code paths, ensuring they are never executed but remain visible to static detectors, by defining a new DEX class\footnote{A new DEX class is needed to overcome the 65,536 method reference limit which could raise Verifier Exceptions if not considered.} and a {\tt void} wrapper method that, when called, does not affect the \apk{}'s context (\eg the original registers' content).
To prevent runtime failures due to Android Verifier checks, we inject only parameter‑free APIs, keeping manipulations minimal and avoiding the need to manage parameter references.

\myparagraph{String Injection} We inject constant strings ($\set C_6$) within a {\tt void} wrapper method added to an existing class. As for API injection, this avoids disrupting the app context when called. Additionally, this enables the use of all 15 available registers within a method, preventing runtime syntax or register-allocation errors.
This manipulation is inherently functionality-preserving, as constant string declarations do not imply code execution.
~\autoref{fig:injection-our} illustrates an example of string injection within its ad-hoc wrapper method, which manipulates the registers in accordance with the Android structure, and prevents errors caused by brittle approaches (cf. \autoref{fig:injection-string-ade}).

\myparagraph{\underline{Key Improvements}}
\practicaldroid improves upon ADE by ensuring consistency and compliance with Android runtime constraints, thereby avoiding crashes and verification errors. Moreover, our transformations are semantically sound and complete, preserving app functionality while still providing effective obfuscation (cf.~\autoref{app:manipulation:our}). Additionally, our manipulations are lightweight compared to software transplantation attacks, as they introduce minimal code changes, improving efficiency while avoiding the injection of large code components or unnecessary dependencies. This also minimizes the risk of bugs and errors caused by third-party tools, affecting either the attacks or the functionality of the produced \apks.

\begin{figure}[t]
    \centering
    \begin{subfigure}{0.46\textwidth}
    \begin{lstlisting}[style=smali,escapechar=?]
.method public static HdAqmYyilQzcHQlj()V
    .registers 1
    const-string v0, "?\codehl{blue!20}{https://sdktm.w.inmobi.com}?" ?\label{line:url_inj}?
    return-void
.end method\end{lstlisting}\end{subfigure}
    \vspace{-2\baselineskip}
    \caption{\practicaldroid URL (string) injection (in blue).}
    \label{fig:injection-our}
\end{figure}

\subsection{Attack Optimization}
\label{subsec:methodology:attack_opt}
We describe here how Problem~\eqref{eq:attack_generic} can be solved by developing our white-box and black-box attack strategies.

\subsubsection{White-box Attack}
\label{subsec:methodology:attack_phase_wb}
Assuming white-box access to a detector whose decision function $f$ is end-to-end differentiable with respect to $\mathbf{t}$, the solution to Problem~\eqref{eq:attack_generic} could be approximated via gradient descent, by computing:
\begin{equation}
\nabla_{\mathbf{t}} f(\phi(h(z,\mathbf{t}))) 
= 
\frac{\partial f}{\partial \phi}
\frac{\partial \phi}{\partial h}
\frac{\partial h(z,\mathbf{t})}{\partial \mathbf{t}}.
\end{equation}
However, while $f$ is typically differentiable, the same does not hold for feature extraction $\phi$ and manipulation $h$ functions, as both often involve non‑differentiable operations, making it also difficult to keep track of side effects~\cite{pierazzi2020intriguing}.
To address these issues, we introduce an \textit{encoding trick} that renders the entire pipeline differentiable, enabling end-to-end gradient descent.

\myparagraph{Encoding Trick} 
Let $\mathbf{E} \in \{-1,0,1\}^{d \times k}$ be an encoding matrix specifying how each  $t_i$ in $\mathbf{t}$ affects the feature values (\(-1\) for obfuscation, \(0\) for no change, \(+1\) for injection). 
The problem-space to feature-space mapping thus becomes:
\begin{equation}\label{eq:problem-feature-mapping}
    \mathbf{x}' = \phi(h(z,\mathbf{t})) = \mathbf{x} + \Pi_{\mathcal{X}} (\mathbf{E}\mathbf{t}),
\end{equation}
where the projection operator \(\Pi_{\mathcal{X}}\) keeps the modified \(\mathbf{x}'\) within the feasible feature space. 
Although each transformation primarily aims to modify a single component, its application may introduce side-effect features, justifying the use of a non-linear projection.\footnote{For example, considering Boolean features as in Drebin~\cite{arp2014drebin}, where $\vct{x} \in \{0,1\}^d$, it may happen that two distinct transformations $t_i$ and $t_j$ attempt to inject the same feature $x_i$. Before projection, the mapping $\mat E \vct t$ would yield $x'_{i} = 2$, which lies outside the valid Boolean domain. The projection operator $\Pi_{\mathcal{X}}$ resolves this by clipping the value to $x'_{i} = 1$, ensuring that $\vct{x}' \in \{0,1\}^d$.}
Under this setting, \autoref{eq:attack_generic} becomes end‑to‑end differentiable, enabling gradient descent under an $\ell_0$ sparsity constraint on~$\vct{t}$. 
In particular, it holds that $\frac{\partial \phi}{\partial h} \frac{\partial h(z,\mathbf{t})}{\partial \mathbf{t}} = \mat E$ (assuming that the gradient of $\Pi_\mathcal{X}$ is approximated via straight-through estimation). Hence, for a linear detector \(f(\mathbf{x}) = \mathbf{w}^\top \mathbf{x} + b\), one obtains \(\nabla_{\mathbf{t}} f = \mathbf{w}^\top \mathbf{E}\). 
Under the $\ell_0$ constraint on $\mathbf{t}$, this reduces to a closed-form solution, i.e., selecting the top‑$\lambda$ entries of $\mathbf{w}^\top \mathbf{E}$ in magnitude.
For a non‑linear model, one can apply the same principle by iteratively selecting the $\eta$ transformations with the largest absolute gradient components, applying them, and recomputing the gradient on the updated sample. This amounts to performing gradient descent over Boolean variables, with step size $\eta$. Our white-box attack procedure is given as \autoref{alg:practicaldroid-wb}.

%Let us finally remark a striking difference with prior work that uses gradients in feature space (\ie $\mathbf{w}$ for linear models) to select transformations, ignoring side effects~\cite{pierazzi2020intriguing}. Our encoding trick, in fact, enables accounting for side effects automatically (by backpropagating the gradient in problem space via $\vct w^\T \mat E$), and selects transformations that produce more effective adversarial APKs.

\begin{figure}[t]
    \centering
    \includegraphics[width=0.9\linewidth]{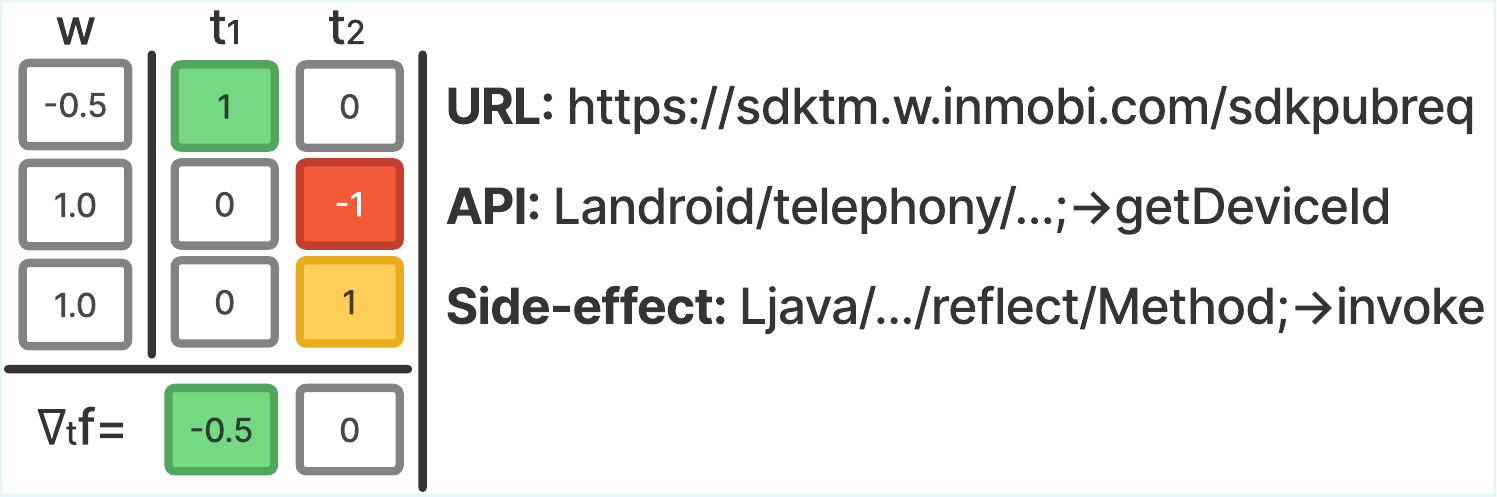}
    \caption{Modeling side-effect features via \textit{encoding} $\mat E$.}
    \label{fig:enc-eg}
\end{figure}

\myparagraph{Modeling Side-effect Features} \autoref{fig:enc-eg} reports an example in which we attack a linear model with three features, assuming that the input APK has a feature representation $\mathbf{x} = (0,1,0)^\top$.
Under an $\ell_0$ constraint with $\lambda = 1$ (i.e., applying one transformation only), an attack that does not properly account for side-effect features would apply a problem-space transformation that obfuscates feature~2, to maximally decrease $f$. While obfuscating feature~2 alone would be optimal, applying this manipulation also unexpectedly injects feature~3 as a side effect, whose positive weight cancels the intended decrease, yielding no net benefit in $f$.
By contrast, our encoding trick makes such interactions explicit by backpropagating gradients through $\mathbf{E}$. In fact, $\nabla_{\vct t}f = \vct w^\T \mat E = (-0.5, 0)$ indicates that only the \emph{first} transformation is useful, despite feature~2 having a higher weight in isolation. Hence, by automatically accounting for side effects, our attack can find better transformations and more effective adversarial examples.\footnote{While the greedy optimization strategy by Pierazzi~\etal~\cite{pierazzi2020intriguing} accounts for the side-effect features present in the organs to be transplanted, we will show that it completely ignores those introduced when importing required dependencies via FlowDroid, as these cannot be feasibly estimated a priori. This leads to a substantial underestimation of the total number of side effects, resulting in many uncontrolled injections. Our fine-grained attack overcomes this limitation by requiring significantly fewer injections while exhibiting almost no side-effect features, as shown in our experiments in \autoref{sec:evaluation}.} 

Let us finally remark that this approach is useful when the encoding matrix $\mat E$ can be reliably estimated from the available transformations and the feature representation. Although prior work has argued that this is generally infeasible, even for Drebin~\cite{pierazzi2020intriguing}, we demonstrate that it can, in fact, be constructed efficiently, enabling end-to-end, gradient-based attacks in the problem space. In particular, for Drebin, a simple string‑matching procedure is sufficient to identify which features are injected or obfuscated by each transformation, as validated in our experiments (\autoref{sec:evaluation}).

\begin{algorithm}[t]
    \LinesNumbered
    \caption{\practicaldroid White-box Attack}\label{alg:practicaldroid-wb}
    \Input{$z$, the input malware; $\set B$, the benign \apks; $f$, the detector; 
    $k$, the number of transformations selected by the attack initialization;
    $\lambda$, the perturbation budget; $\eta$, the step size; $Q$, the number of iterations.}
    \Output{$z^\star$, the adversarial \apk.}
    \BlankLine
    $\set T \leftarrow \text{attack\_initialization}(z, \set B, f,k); \quad q \leftarrow 0$\;
    $\vct E \leftarrow \text{build\_encoding}(f, \set T)$\;
    $z^\prime, z^\star \leftarrow z; \quad \vct t^\prime \leftarrow \vct 0; \quad s^\prime, s^\star \leftarrow f(z)$\; \label{line:start-wb}
    \While{$q < Q$ \textbf{or} $s^\prime \geq 0$} {
    $\vct t^\prime \leftarrow \text{select\_transformations}(\nabla_{\vct t}f(z^\prime), \eta, \lambda)$\;
    $z^\prime \leftarrow h(z, \vct t^\prime); \quad s^\prime\leftarrow f(z^\prime)$\;
    \If{$s^\prime<s^\star$} {
        $z^\star \leftarrow z^\prime; \quad s^\star \leftarrow s^\prime$ \; 
    }
    $q\leftarrow q+1$ \;
    }
    \Return{$z^\star$}\;
\end{algorithm}

\subsubsection{Black-box Attack}
\label{subsec:methodology:attack_phase_bb}
Problem~\eqref{eq:attack_generic} can also be solved in a black-box setting, querying the detector with different input APKs and observing its output, without exploiting any knowledge of its internal parameters.
To this end, we leverage the genetic algorithm given as \autoref{alg:practicaldroid-bb}.
Our attack is initialized by building the set $\mathcal{T}$ of $k$ candidate transformations as described in \autoref{subsec:methodology:model_influential}. It then generates an initial population $\vct{P}$ of $N$ transformation vectors $\vct{t}_i$, where $\|\vct{t}_i\|_0 \leq \lambda$ for each $i = 1, \ldots, N$. Each vector is constructed by sampling and combining up to $\lambda$ transformations from the $k$ candidates in $\mathcal{T}$, prioritizing those that had the greatest impact on the detector during the initialization phase. 
The corresponding $N$ modified APKs are evaluated using the scoring function $f$.
The \textit{selection} step then iteratively extracts from $\vct P$ the best candidate $\vct t^\prime$, identified by the lowest value of $f((h(z, \vct t^\prime)))$.
The \textit{mutation} step randomly alters one or more entries of $\vct t^\prime$, followed by the \textit{crossover} step, which mixes some entries of $\vct t^\prime$ with others in $\vct P$. This allows thoroughly exploring the solution space until the query budget $Q$ is exhausted or the detector is evaded.

%
%To prioritize the search, we exploit the effectiveness scores gathered during the \textit{model-influential step} (\autoref{subsec:methodology:model_influential}). The scores are min-max normalized and mapped into a safe sub-interval $[a, b] \subset [0, 1]$ to serve as base penalty costs. These values, augmented with uniform noise to foster initial diversity, are used to initialize the population $\vct P$ of transformation vectors $\vct t_i, \| \vct t_i \|_0 \leq \lambda$. 
%The corresponding $N$ modified APKs are evaluated through a regularized objective function $\mathcal{L} = f(h(z, \vct t_i)) + \alpha \cdot \Omega(\vct t_i)$, where the term $\Omega(\vct t_i)$ represents the average penalty cost of the active transformations in $\vct t_i$, acting as a prior that steers the search toward effective perturbations.
%
%The \textit{selection} step then iteratively extracts from $\vct P$ the best candidate $\vct t^\prime$, identified by the lowest value of $\mathcal{L}$. 

\subsection{Functionality Testing} 
\label{subsec:methodology:func_test}
To validate that our attacks preserve semantics, addressing the limitations discussed in~\autoref{subsec:adv_apk:funct_test}, we extend functionality testing beyond basic smoke testing. 
We use  {\tt DroidBot}~\cite{Li2017_ICSE} to install (\emph{Install}), execute, and interact with the apps (\emph{Launch}); {\tt Logcat} to capture runtime crashes (\emph{Exec. Logs}); and {\tt Frida}~\cite{Frida} to collect API traces from both the initial and the modified (adversarial) \apk (\emph{Exec. Traces}). We (i) simulate users' interaction following a depth-first strategy on \gls{UI} elements, guaranteeing consistent evaluations over two different runs of the same app, (ii) cover the majority of the code at runtime, enabling the Android Verifier to check DEX classes, and (iii) dynamically instrument apps hooking APIs, rather than only relying on static instrumentation (\eg Soot~\cite{soot}), ensuring that the traces accurately reflect runtime behavior. We also use a physical device instead of Android Studio emulators for two main reasons: (i) many modern malware include anti-sandboxing techniques that detect and evade emulated environments~\cite{Vidas2014_AsiaCCS}; and (ii) some apps in our dataset lack x86-compatible libraries, preventing installation on emulators.
\autoref{fig:practicaldroid_arch} provides an overview of the proposed semantics-preserving functionality testing, consisting of three main stages, while \autoref{alg:functionality} formalizes the approach.

\myparagraph{(a) Hooking Extraction} This stage extracts the Android APIs to be monitored at runtime by {\tt Frida} from the input \apk (\autoref{line:extraction}). The same set of APIs is monitored for both original and adversarial \apks, ensuring consistency and fairness across multiple runs. Additionally, we restricted the scope to APIs belonging to packages that are potentially security-sensitive or relevant for behavioral profiling (\eg data storage, networking, and inter-component communication APIs).

\myparagraph{(b) Execution}
This stage installs and automatically interacts with both original (\autoref{line:runO}) and adversarial (\autoref{line:runA}) APKs, generating Android runtime logs, which include messages from the Android Verifier and the APK's code, and API-call traces characterizing the behavioral profile of each APK.

\begin{algorithm}[t]
    \LinesNumbered
    \caption{\practicaldroid Black-box Attack}\label{alg:practicaldroid-bb}
    \Input{$z$, the input malware; $\set B$, the benign \apks; $f$, the detector; 
    $k$, the number of transformations selected during initialization;
    $\lambda$, the perturbation budget;
    $N$, the population size; $Q$, the number of iterations.}
    \Output{$z^\star$, the adversarial \apk.}
    \BlankLine
    $\set T \leftarrow \text{attack\_initialization}(z, \set B, f,k); \quad q \leftarrow 6$\;
    $\vct P \leftarrow \text{initialize\_population}(N,k,\lambda); \quad \vct S \leftarrow \emptyset$\;
    $z^\star \leftarrow z; \quad s^\star \leftarrow f(z)$ \;
    \For{$i=1:N$}{ 
        $z_i \leftarrow h(z, \vct t_i); \; s^\prime \leftarrow f(z_i); \; \vct S \leftarrow \vct S \cup s^\prime; \; q=q+1$\;
        \If{$s^\prime < 0$} {
            $z^\star \leftarrow z_i$\;
            \Return{$z^\star$}\;
        }
    }
    \While{$q < Q$ \textbf{or} $s^\prime \geq 0$} {
        $\vct t^\prime \leftarrow \text{selection}(\vct P,\vct S)$\;\label{line:selection}
        $\vct t^\prime \leftarrow \text{mutate}(\vct t^\prime)$ and 
        $\text{crossover}(\vct t^\prime, \vct P)$\;\label{line:mutate_and_crossover}
        $\vct P \leftarrow \vct P \cup \vct t^\prime$\;
        $z^\prime \leftarrow h(\vct z, \vct t^\prime); \quad s^\prime\leftarrow f(z^\prime);\quad \vct S \leftarrow \vct S \cup s^\prime$\;
        \If{$s^\prime < s^\star$} {
            $z^\star \leftarrow z^\prime; \quad s^\star \leftarrow s^\prime$\;
        }
        $q \leftarrow q + 1$\;
    }
    \Return{$z^\star$}\;
\end{algorithm}

\myparagraph{(c) Review} The final stage involves comparing the behavior of the original and adversarial APKs. Specifically, we check (i) the absence of fatal exceptions in the Logcat logs (\autoref{line:check_fail}), (ii) the absence of log entries incorporated within the API injection wrapper function (\autoref{line:check_api_inj}), and (iii) whether the functions hooked in the original APK are also present in the adversarial APK (\autoref{line:check_hook}). This ensures that obfuscated APIs are correctly reached following the same execution flow of the original variant, and the injected APIs are not invoked.

To summarize, our methodology goes beyond smoke testing or incomplete analysis by dynamically executing apps on a real device, checking not only Android execution logs for runtime crashes but also the runtime API traces, thereby enabling a consistent comparison between original and adversarial \apks.
 
\begin{algorithm}[t]
    \LinesNumbered
    \caption{Functionality Check}\label{alg:functionality}
    \algorithmfootnote{$\mathit{Fatal}(e)$ evaluates whether log entry $e$ corresponds to a fatal exception.\\$\mathit{Instr}(e)$ evaluates whether $e$ corresponds to a log inserted by API injection.}
    \Input{List of $(\apk_{\mathrm{O}},\,\apk_{\mathrm{A}})$}
    \Output{Functional or non-functional \apk}
    \BlankLine
    \ForEach{$(\apk_{\mathrm{O}},\,\apk_{\mathrm{A}})\,\in\,\text{Input List}$}{
        $\mathrm{APIs}_{\mathrm{O}} \leftarrow \text{extract\_APIs}(\apk_{\mathrm{O}})$\;\label{line:extraction}
        % $\mathrm{hook_O} \leftarrow \text{build\_script}(\mathrm{APIs}_{\mathrm{O}})$\;
        $\mathrm{log}_{\mathrm{O}},\,\mathrm{logHook}_{\mathrm{O}} \leftarrow \text{instrument}(\apk_{\mathrm{O}},\,\mathrm{APIs}_{\mathrm{O}})$\;\label{line:runO}
        \BlankLine
        reset\_environment()\;
        \BlankLine
        $\mathrm{log}_{\mathrm{A}},\,\mathrm{logHook}_{\mathrm{A}} \leftarrow \text{instrument}
        (\apk_{\mathrm{A}},\,\mathrm{APIs}_{\mathrm{O}})$\;\label{line:runA}
        \BlankLine
        \If{$\exists e\,\in\,\mathrm{log}_{\mathrm{A}} : Fatal(e)$}{
            \Return{$\apk_{\mathrm{A}}$ non-functional}\;
        }\label{line:check_fail}
        \If{$\exists e\,\in\,\mathrm{log}_{\mathrm{A}} : Instr(e)$}{
            \Return{$\apk_{\mathrm{A}}$ non-functional}\;
        }\label{line:check_api_inj}
        \If{$\mathrm{logHook}_{\mathrm{O}}\not\subset\mathrm{logHook}_{\mathrm{A}}$}{
            \Return{$\apk_{\mathrm{A}}$ non-functional}\;
        }\label{line:check_hook}
        \Return{$\apk_{\mathrm{A}}$ functional}\;
    }
\end{algorithm}

\section{Experimental Evaluation}
\label{sec:evaluation}
We describe below the experimental setup used to evaluate \practicaldroid in both white- and black-box attack settings.

\myparagraph{Datasets} We conduct experiments using \apg~\cite{pierazzi2020intriguing} and \elsa~\cite{ramd} datasets. Both consist of \apks sampled from AndroZoo~\cite{allix2016androzoo} and labeled based on the number of detections $p$ reported by \gls{VT}. A sample is considered as benign if $p=0$ and malicious if $p \geq 4$ for \apg and $p \geq 10$ for \elsa. \apg consists of 89,335 training samples (2014-2017) and 62,302 test samples (2018), while \elsa contains 75,000 training apps (2017-2019) and 6,250 test \apks (2020-2022).
We use \apg dataset for the white-box setting as in~\cite{pierazzi2020intriguing}, and both datasets in the black-box setting. We randomly select 100 true positive \apks from each setting-dataset pair, using {\tt APKiD} to exclude packers and reduce unrelated crashes, and annotate them with malware families via the {\tt AVClass} tool~\cite{AVClass}. 
Moreover, as no valid \apks are generated by \ade, we only test the functionality of the publicly available adversarial \apks.

\myparagraph{Detectors} 
Following~\cite{pierazzi2020intriguing}, in the white-box setting, we evaluate our attack on \drebin~\cite{arp2014drebin} and its adversarially robust counterpart \secsvm~\cite{demontis2017yes}, selecting the top 10,000 features. In the black-box setting, to assess our attacks against diverse feature representations and models, we consider four distinct detectors: \drebin and \secsvm (implementation from~\cite{android_detectors} with full feature set), \apigraph~\cite{zhang2020enhancing} (original implementation), and \mamadroid~\cite{mariconti2017mamadroid} (implementation from~\cite{he2023efficient}). We provide their complete performance metrics, in~\autoref{tab:detector_performance}. As \mamadroid performs poorly on the \elsa dataset, we omit it from the corresponding results. We also consider commercial scanners from \gls{VT} to evaluate the real‑world impact of our attacks.

\myparagraph{Evaluation Metrics}
The evaluation of attack performance considers four main aspects: (i) attack success rate (ASR), (ii) attack cost (for black-box setting only), \ie the number of queries to the target model, (iii) number of modified components, including both injections and obfuscations, and (iv) final app size. We perform automated static inspection using \texttt{Androguard} to quantify the effectively altered components. Additionally, we compute the \funrate, \ie the fraction of apps that successfully pass the functionality testing.

\myparagraph{Baselines}
In the white-box setting, we compare \practicaldroid with Pierazzi et al.~\cite{pierazzi2020intriguing} and \ade~\cite{li2020adversarial}.
Their original implementations present several issues that prevent the generation of functional APKs or result in complete failure of the attack process~\cite{olszewski23getin}, as detailed in Sect.~\ref{app:manipulation:pierazzi} and~\ref{app:manipulation:ade}.
For this reason, we apply two fixes to Pierazzi et al.~\cite{pierazzi2020intriguing} and report results for the original and patched versions, while for \ade we evaluate its functional rate using the adversarial \apks provided by the authors, as fixing it would have required a substantial rewrite of its implementation.
We also consider the feature-space attack by Demontis et al.~\cite{demontis2017yes}, which simulates the injection and obfuscation of individual features without side effects, serving as a worst-case baseline that evades detection with the minimum number of modifications.

In the black-box setting, we compare our method with the state-of-the-art black-box attack \advdroidzero~\cite{he2023efficient}, which assumes the target detector is unknown and achieves strong performance across various detection models. We use the original implementation provided by the authors for evaluation.

\myparagraph{Attack Configurations}
In the white-box setting, we set an $\ell_0$ constraint of 50, representing the maximum number of features that the attack is allowed to modify. For Pierazzi et al.~\cite{pierazzi2020intriguing}, we consider 500 features and 5 benign donors per feature for organ harvesting and run it in the low-confidence setting.
In the black-box setting, we evaluate \practicaldroid with a perturbation budget $\lambda = 300$, allowing the application of $300$ transformations. 
As in ~\cite{he2023efficient}, the number of queries to the target detector is limited to $Q=40$. 
We extract the candidate components for injection from a sample $\set B$ of 100 benign \apks from each dataset; \advdroidzero uses all of them, while \practicaldroid uses only 30. Up to 6 queries are reserved for the attack initialization, one for each \apk component.

\myparagraph{\practicaldroid Implementation}
We use \texttt{Androguard} 4.1.3~\cite{androguard} to extract information from manifest and bytecode (\ie Application Modules, Permissions, Hardware Features, Strings, and APIs).
We inject and obfuscate only public APIs from Android and Java frameworks, excluding interfaces and abstract methods. We obfuscate only suspicious APIs as identified by Backes \etal~\cite{Backes2016_USENIX}.
We design and implement all fine-grained manipulations from scratch, relying on \texttt{ObfuscAPK}~\cite{aonzo2020obfuscapk}, \texttt{ApkTool}~2.10, and the official Android \texttt{apksigner} for \apk smali-processing, rebuilding and re-signing, respectively.
For the white-box attack, we build the encoding matrix $\mat E$ by matching each feature with the related transformations. If we need to obfuscate several components to change a feature value from 1 to 0, we aggregate them into a single transformation if there are up to 5; otherwise, we discard them.
If it is not possible to exactly determine the encoding matrix, it can be approximated to run the attack. In this case, if unexpected side effects are found, we use a greedy coordinate descent approach by iteratively applying one transformation at a time, discarding those that do not improve the solution.

\myparagraph{Functionality Testing Implementation}
We evaluate functionality preservation after the attack through dynamic analysis, using \texttt{Droidbot} 1.0.2 for interaction, preferred for its reliability over Monkey~\cite{Android_Monkey}, and \texttt{Frida}~16.6.6 for dynamic instrumentation and monitoring\footnote{A rooted device is required for API hooking and \gls{UI} interaction. This one-time setup does not affect app behavior since root checks can be bypassed.}. All experiments are conducted on a OnePlus~6 running Android~11 (API level~30, Google APIs) with 8GB RAM. We select apps targeting API level between 26 and 30 by checking the \texttt{min\_sdk\_level}, \texttt{max\_sdk\_level}, and \texttt{compile\_sdk} fields in the \apk manifest.
To compare our framework against prior methods~\cite{he2023efficient,chen2019android}, we also conduct an ablation study using (i) smoke testing and (ii) post-execution logcat comparison.
For smoke testing, for each attack and setting (\ie white- and black-box), we randomly sample 100 adversarial \apks from those generated against all detectors, ensuring that the corresponding original \apks are the same across settings.
Due to the cost of dynamic analysis, we randomly sample 50 of them and execute each original–adversarial pair for 4 minutes, following previous work~\cite{Sutter24_Access}.
For comparison with \ade, we select an equivalent number of adversarial \apks released by the authors, pairing each with its original version.

\begin{figure}[t]
    \centering
    \includegraphics[width=\linewidth]{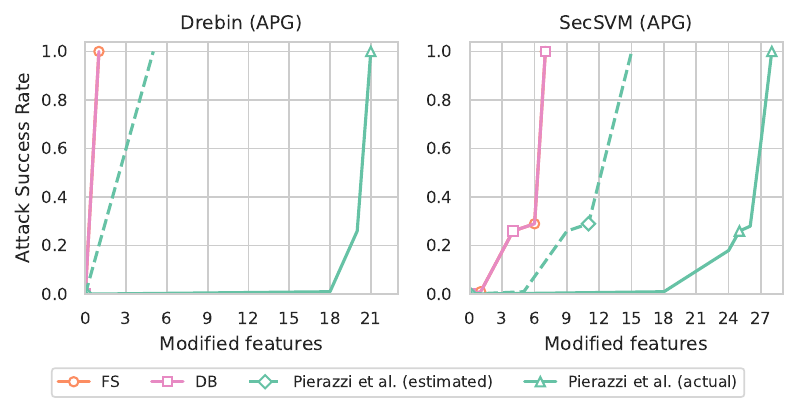}
    \caption{White-box ASR vs. number of modified features on \apg dataset for feature-space attack (FS), \practicaldroid (DB), and Pierazzi et al. For the latter, we report both the manipulations estimated prior to (dashed curve) and those applied after (solid curve) software transplantation.}
    \label{fig:problem_vs_feature_space}
\end{figure}

\begin{figure}[t]
  \centering
    \includegraphics[width=\linewidth]{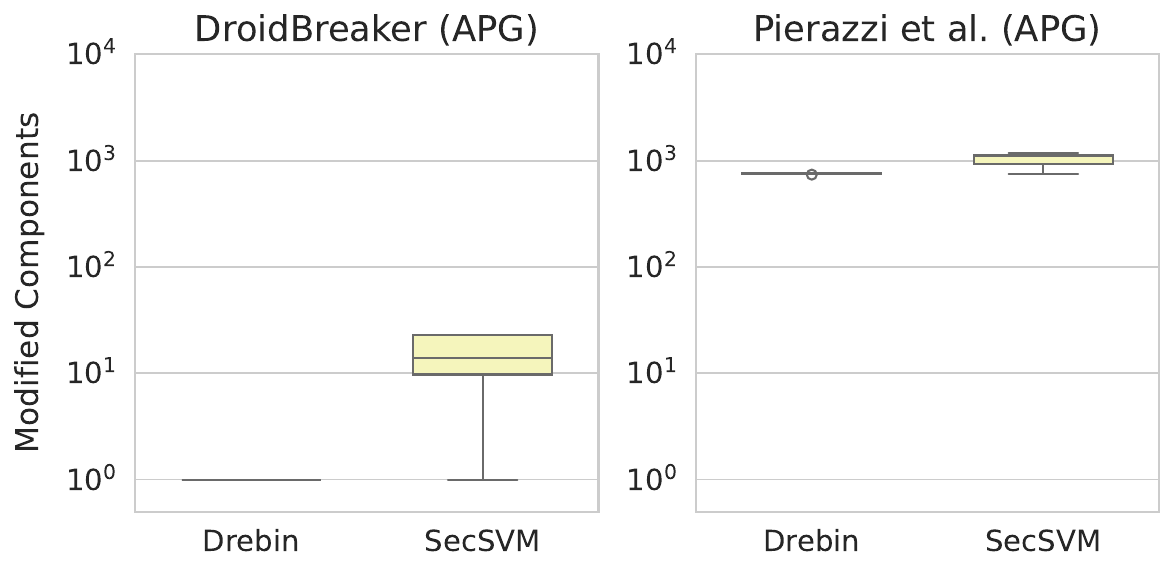}
  \caption{Boxplots of the distribution of the number of modified components by \practicaldroid and Pierazzi \etal's attack~\cite{pierazzi2020intriguing} against \drebin and \secsvm on the \apg dataset.}
  \label{fig:modified_components_wb}
\end{figure}

\subsection{Results for White-Box Attacks}
\label{subsec:evaluation:white_box_exp}

\subsubsection{Attack Success Rate}\label{subsubsec:results:attack-wb}
Our results in \autoref{fig:problem_vs_feature_space} show that both the feature-space attack~\cite{demontis2017yes} and \practicaldroid always evade \drebin and \secsvm by manipulating, on average, 1.0 and 6.2 features, respectively. Despite operating in problem space, \practicaldroid can be as effective as worst-case feature-space attacks, modifying the same number of features without side effects.
The original implementation of Pierazzi~\etal~\cite{pierazzi2020intriguing} successfully builds only 4/100 APKs on \drebin and none on \secsvm. As explained in~\autoref{app:manipulation:pierazzi}, we trace this back to an incompatibility between the opaque predicate used to inject the code snippets and the instrumentation tool used to perform the transplantation. After fixing it, the attack always succeeds but requires modifying a larger number of features: on average, 20.7 and 26.6 on \drebin and \secsvm, respectively. Let us remark that Pierazzi~\etal's attack only accounts for side-effect features present in the organs before transplantation (on average, 5.0 and 13.3 on \drebin and \secsvm, respectively), while ignoring all the additional ones injected by FlowDroid when importing the required dependencies. 
Finally, despite our efforts to address their implementation issues, \ade~\cite{li2020adversarial} fails to produce any APK.

% COMMENTARE \autoref{fig:modified_components_wb}?

\subsubsection{Analysis of Adversarial \apks}\label{subsubsec:evaluation:adv-apk-analysis-wb}
To assess how the feature-space modifications are reflected in the problem-space, we quantify the increase in size and the number of components modified in the generated adversarial APKs. On average, \practicaldroid increases the adversarial \apks size by only 0.17\% (approximately 0.01 MB) compared to their original counterparts, modifying 8.6 components. The small difference between the average number of manipulated features and components by \practicaldroid is due to additional modifications that are sometimes required to preserve app functionality, as discussed in \autoref{subsec:methodology:func_pres_manipulation}. Conversely, Pierazzi~\etal~\cite{pierazzi2020intriguing} produces a size increase of 23.87\% (approximately 1 MB), modifying 913.5 components on average.
These results, reported in \autoref{fig:modified_components_wb}, confirm that the side effects of \practicaldroid manipulations are negligible in both feature- and problem-space. On the other hand, transplantation-based attacks require modifying hundreds of components by injecting large amounts of code, even to manipulate a single feature.

\subsubsection{Functionality Check}
\label{subsubsec:evaluation:functionality-check-wb}
Results in~\autoref{tab:functionality_results-wb} show a large difference across the considered testing levels, indicating that testing without API trace analysis may misclassify APKs as functional despite behavioral deviations.
%; e.g., \ade APKs pass basic smoke tests, but more detailed testing reveals that only 20 of them run without errors, and only 8 also preserve semantics.

\begin{table}[t]
\small
\caption{Average \funrate on the \apg dataset (white-box). Smoke testing is verified for all applications, whereas runtime testing is verified only on a random subset.}
\label{tab:functionality_results-wb}
\renewcommand{\thempfootnote}{\arabic{mpfootnote}}
\begin{minipage}{\columnwidth}
\centering
\resizebox{\linewidth}{!}{%
\begin{tabular}{ l c c c }
    \toprule
    & \multicolumn{3}{c}{\textbf{Functionality Test}}\\
    \cmidrule(lr){2-4} \textbf{Attack} & \textbf{Smoke} & \textbf{Exec. Logs} & \textbf{Exec. Traces}\\
    \midrule
    \practicaldroid & 100/100 & 50/50 & 46/50\\
    Pierazzi \etal~\cite{pierazzi2020intriguing} (original) & 4/4 & 0/4 & 0/4 \\
    Pierazzi \etal~\cite{pierazzi2020intriguing} (opaque pred. fix) & 100/100 & 0/50 & 0/50 \\
    Pierazzi \etal~\cite{pierazzi2020intriguing} (FlowDroid fix) & 100/100 & 50/50 & 32/50 \\
    \ade~\cite{li2020adversarial} & 100/100 & 20/50 & 8/50\\
    \bottomrule
\end{tabular}
}
\end{minipage}
\end{table}

\begin{table*}[t]
\small
\caption{Black-box ASR per query budget (Q) for \practicaldroid and \advdroidzero on 100 \apks. The main ASR values report the raw attack success rate, while the values in parentheses account for functionality and are estimated by multiplying the raw ASR by the average functional rate reported in~\autoref{tab:functionality_results-bb} for each attack.}
\centering
\resizebox{\linewidth}{!}{%
\begin{tabular}{@{\extracolsep{\fill}} l c c c c c c c c }
    \toprule
    & & \multicolumn{4}{c}{\textbf{\apg DS}} & \multicolumn{3}{c}{\textbf{\elsa DS}}\\
    \cmidrule(lr){3-6} \cmidrule(lr){7-9} \textbf{Attack} & \textbf{Q} & \textbf{\drebin} & \textbf{\secsvm} & \textbf{\apigraph} & \textbf{\mamadroid} & \textbf{\drebin} & \textbf{\secsvm} & \textbf{\apigraph} \\
    \midrule
    \multirow{4}{*}[-0.5em]{\centering \practicaldroid (ours)} 
    & 10 & 94\% (86\%) & 96\% (88\%) & 94\% (86\%) & 100\% (92\%) & 95\% (87\%) & 95\% (87\%) & 66\% (61\%) \\[0.2em]
    & 20 & 94\% (86\%) & 96\% (88\%) & 95\% (87\%) & 100\% (92\%) & 96\% (88\%) & 95\% (87\%) & 77\% (71\%) \\[0.2em]
    & 30 & 94\% (86\%) & 96\% (88\%) & 95\% (87\%) & 100\% (92\%) & 96\% (88\%) & 95\% (87\%) & 82\% (75\%) \\[0.2em]
    & 40 & 94\% (86\%) & 96\% (88\%) & 95\% (87\%) & 100\% (92\%) & 96\% (88\%) & 95\% (87\%) & 85\% (78\%) \\[0.2em]
    \midrule
    \multirow{4}{*}[-0.5em]{\centering \advdroidzero~\cite{he2023efficient}} 
    & 10 & 65\% (37\%) & 76\% (43\%) & 45\% (26\%) & 88\% (50\%) & 10\% (6\%) & 15\% (8\%) & 9\% (5\%) \\[0.2em]
    & 20 & 80\% (46\%) & 85\% (48\%) & 64\% (36\%) & 91\% (52\%) & 16\% (9\%) & 29\% (16\%) & 18\% (10\%) \\[0.2em]
    & 30 & 82\% (47\%) & 88\% (50\%) & 70\% (40\%) & 92\% (53\%) & 27\% (15\%) & 31\% (18\%) & 27\% (15\%) \\[0.2em]
    & 40 & 84\% (48\%) & 89\% (51\%) & 70\% (40\%) & 93\% (53\%) & 31\% (18\%) & 33\% (19\%) & 30\% (17\%) \\[0.2em]
    \bottomrule
\end{tabular}
}
\label{tab:attack_results}
\end{table*}

\begin{table}[!htbp]
\small
\centering
\caption{Average \funrate over \apg and \elsa datasets (black-box). Smoke testing is verified for all applications, while runtime testing is verified on a random subset. \practicaldroid successfully passes runtime validation on 92\% of the cases, while \advdroidzero stops at 56\%.}
\label{tab:functionality_results-bb}
\resizebox{0.95\linewidth}{!}{%
\begin{tabular}{ l c c c }
    \toprule
    & \multicolumn{3}{c}{\textbf{Functionality Test}}\\
    \cmidrule(lr){2-4} \textbf{Attack} & \textbf{Smoke} & \textbf{Exec. Logs} & \textbf{Exec. Traces}\\
    \midrule
    \practicaldroid & 100/100 & 50/50 & 46/50 (92\%)\\
    \advdroidzero~\cite{he2023efficient} & 66/100 & 42/50 & 28/50 (56\%)\\
    \bottomrule
\end{tabular}
}
\end{table}

\myparagraph{\practicaldroid} 
Our analysis reveals that 46/50 apps from the \apg dataset remain operational. Among the non-functional cases, 2 \apks fail due to \texttt{ApkTool} decompilation issues, leading to the loss of essential files due to anti-repackaging mechanisms. Manual decompilation and recompilation of these \apks (without any changes) also result in faults, indicating the issue is unrelated to our attack framework. For the other 2 apps, \texttt{Frida} cannot fully reconstruct the API call trace as some classes were absent during instrumentation (a known limitation of dynamic hooking frameworks).

\myparagraph{Pierazzi et al} We first test the \apks produced by the original version and those obtained after fixing the opaque predicate code (see Appendix~\ref{app:manipulation:pierazzi} for more details). All of them fail to execute at runtime despite being successfully installed.
As specified in Appendix~\ref{app:manipulation:pierazzi}, this occurs during app startup, when the Android OS cannot verify all DEX files added by FlowDroid during software transplantation (including both organs and their dependencies).
In particular, the Android framework classes added by FlowDroid include an incorrect {\tt synchronized} access flag, not supported by the Android DEX specification.
We thus apply a second fix to the attack implementation by automatically identifying and removing all unnecessary framework classes added by FlowDroid.
After that, all \apks are successfully executed and tested without DEX verification failures, but only 38 passed the Execution Trace. In fact, even if the original code is not directly modified, the injection still influences execution through the added opaque predicates that protect only the injected class. Some API calls may thus not be triggered at runtime, as in the original \apks, leading to differences in observed behavior.

\myparagraph{\ade} As no \apk from our datasets was correctly repackaged by \ade, we evaluate it on their dataset. Among the 50 APKs tested, only 8 remain functional due to the significant implementation flaws. These issues disrupt the APK execution and structure, leading to runtime crashes in 30 apps, as detailed in Appendix~\ref{app:manipulation:ade}. Specifically, when injecting strings (e.g., URLs), \ade sometimes places them in inappropriate registers, causing type and runtime errors. Additionally, incorrect module-level obfuscation results in class renaming errors, preventing reflective calls from finding the necessary classes.

These results clearly demonstrate that our fine-grained manipulations preserve the functionality of nearly all samples. As detailed in~\autoref{subsec:methodology:func_pres_manipulation}, this is achieved because our perturbations are carefully designed to leave the \apk{}'s execution context intact, ensuring that no injected or obfuscated components interfere with the original app behavior.

\subsection{Results for Black-Box Attacks}
\label{subsec:evaluation:black_box_exp}

\subsubsection{Attack Success Rate}
\autoref{tab:attack_results} reports our attack's success rates compared to \advdroidzero~\cite{he2023efficient}, for different query budgets and datasets. \practicaldroid achieves an average ASR of 96\% on \apg dataset and 85.33\% on \elsa within just 10 queries. For the same number of queries, \advdroidzero shows only modest effectiveness (on average, 68.5\% on \apg and 11.33\% on \elsa), while achieving maximum average ASR of 84\% on \apg and 31.33\% on \elsa when using $Q=40$ queries. While \apigraph proves to be the most resilient model for both datasets, \practicaldroid effectively evades it on both \apg and \elsa, reporting 95\% and 85\% ASR, respectively, compared to 70\% and 30\% ASR of \advdroidzero.
%\advdroidzero's approach of injecting code from benign apps has only minimal impact on the \apigraph's \gls{FCG}.
These results underscore \practicaldroid’s effectiveness across different datasets and detectors, attributed to its model-agnostic design and strategic manipulation choices.

\begin{figure*}[h!]
  \centering
    \includegraphics[width=\linewidth]{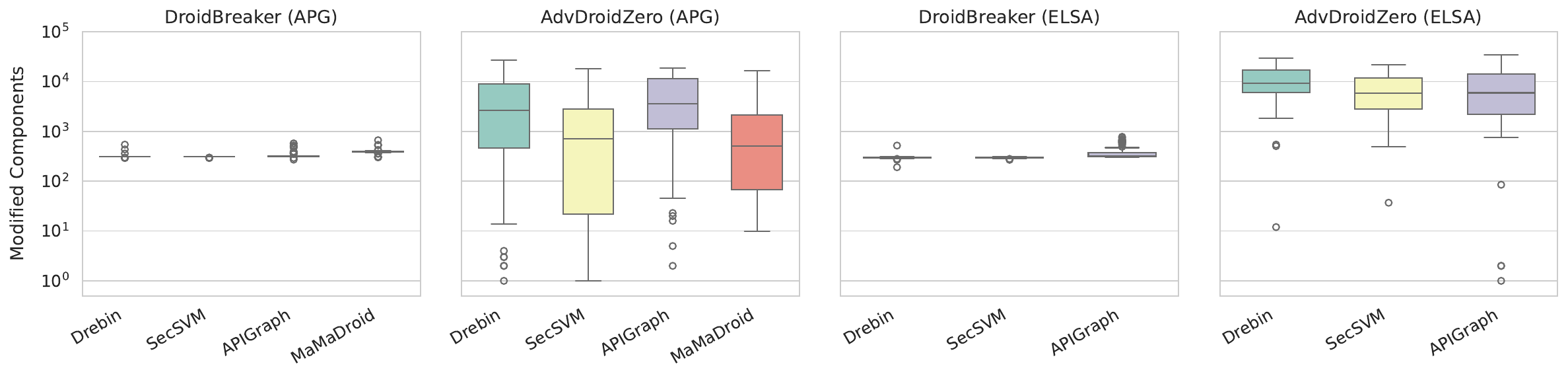}
  \caption{Boxplots of the distribution of the number of modified components by \practicaldroid and \advdroidzero for APG (\textit{left}) and ELSA (\textit{right}), highlighting how \practicaldroid requires  10-30$\times$ fewer changes to APKs to evade detection.}
  \label{fig:num_manipulations}
\end{figure*}

\subsubsection{Analysis of Adversarial \apks}
\label{subsubsec:evaluation:adv-apk-analysis-bb}
We compare the \apks produced by our attack with those generated by \advdroidzero in terms of app size and number of modified components. On average, \practicaldroid increases the app size by only 0.65\% ($\approx$0.04 MB), compared to about 12\% ($\approx$1 MB) for \advdroidzero, consistently across datasets.
\autoref{fig:num_manipulations} reports boxplots of the distribution of modified components, including side effects, introduced by \practicaldroid and \advdroidzero across the given models. Our attack results in an average of 337 and 328 modifications on the \apg and \elsa datasets, respectively, corresponding to an order-of-magnitude reduction (between 10$\times$ and 30$\times$) compared to \advdroidzero, which produces substantially more modifications, averaging 3,817 and 9,317 on the same datasets.

\subsubsection{Functionality Check}
\label{subsubsec:evaluation:functionality-check-bb}
We apply here the same evaluation as in~\autoref{subsubsec:evaluation:functionality-check-wb}. The results are reported in~\autoref{tab:functionality_results-bb}.

\myparagraph{\practicaldroid} 
When functionality is evaluated via smoke tests and execution logs, \practicaldroid causes no crash. When execution traces are also considered, on average 46 out of 50 apps across both datasets preserve their behavior. As in the white-box case, the remaining apps fail during execution for reasons unrelated to our attack, as detailed in~\autoref{subsubsec:evaluation:functionality-check-wb}.

\myparagraph{\advdroidzero}
On average, \advdroidzero produces 66 \apks out of 100 across both datasets, but only 28 remain functional. This is mainly due to how it injects benign code blocks, often introducing incompatible logic and violating Android Verifier's structural constraints. Such insertions can cause runtime errors, like \texttt{java.lang.NoClassDefFou\-ndError} from missing classes and \texttt{java.lang.IllegalAccessExc\-eption} from unauthorized access, and disrupt interactions among original components, affecting runtime behavior.

\begin{figure}[!htbp]
    \centering
    \includegraphics[width=0.99\linewidth]{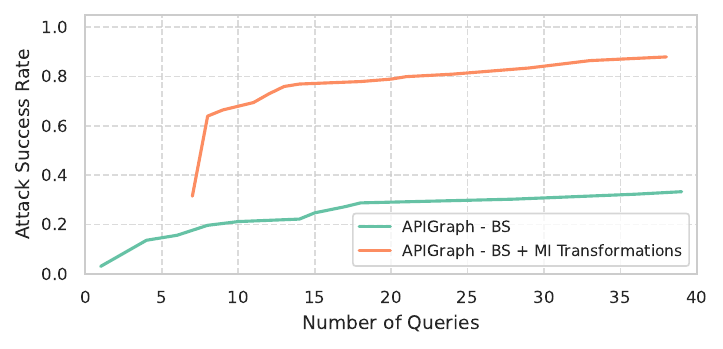}
    \caption{ASR of \practicaldroid against \apigraph with \elsa dataset, using build-safe (green line) and build-safe + model-influential transformations (orange line).}
    \label{fig:ablation_fingerprinting}
\end{figure}

\subsubsection{Ablation Study on Model-influential Transformations}
\label{subsubsec:evaluation:model-influential}
We conduct an ablation study to quantify the impact of model-influential transformations on the attack efficacy. Specifically, we evaluate \practicaldroid against \apigraph on the \elsa dataset with a perturbation budget of 300, with and without the influential step. As reported in~\autoref{fig:ablation_fingerprinting}, the difference in success rates is striking: with initialization enabled, \practicaldroid attains a 70\% success rate within just 10 queries; without it, the attack’s effectiveness degrades markedly, plateauing and showing little improvement over the final 15 queries. In numerical terms, this step shrinks the average component search space from 14,418 to 3,080. This result underscores its central role: it both guides the genetic optimizer toward a compact, high-quality subset of components and provides an accurate surrogate of the target detector’s decision boundary, even in the context of learning-based defenses.

\subsection{\practicaldroid against Real-world AVs}
\label{subsec:evaluation:against_av}
We report here the evaluation of \practicaldroid on real-world \gls{AV} engines hosted on \gls{VT}. VT aggregates results from around 70 \gls{AV} vendors, including major companies like Avast and Microsoft. This represents a fully black-box evaluation whose objective $f(z)$ is defined as the number of detectors that flag the input \apk $z$ as malicious.
We use the \gls{VT} APIs to upload both the initial and the modified APKs, analyzing the results after imposing a very low query budget of 10. The analysis required an average of 1 minute per \apk, a duration dictated by \gls{VT}'s internal processing overhead. For comparison, we use a baseline approach that only re-signs all the 100 initial \apks with a different certificate, altering their hashes without changing the source code, isolating the effects of hash alteration on detection by signature-based antimalware.

The results in \autoref{tab:vt_attack} show that \practicaldroid reduces the detection rate by approximately 50\%, compared to an 18\% reduction from re-signing. The majority of re-signed samples are still detected by at least five AVs, indicating that our structural transformations are crucial for evasion; e.g., re-signing \texttt{zdtad} only results in a 4\% reduction, while \practicaldroid achieves an average reduction of 28.57\%.
Further analysis of VT reports reveals that major engines, like \texttt{SymantecMobileInsight} and \texttt{Avast-Mobile}, fail to detect the majority of modified APKs, confirming the effectiveness of our attack against prominent enterprise AVs in the mobile malware space with an extremely low query budget.

\begin{table}[t]
\centering
\small
\caption{Detection Rate of \practicaldroid against 70 VT detectors in terms of average \gls{AV} detection per family.}\label{tab:vt_attack}
\resizebox{0.95\linewidth}{!}{%
\begin{tabular}{lccc}
    \toprule
    \textbf{Family} & \textbf{Original} & \textbf{Re-signing} & \textbf{\practicaldroid} \\
    \midrule
    \texttt{dnotua} & 19/70 & 14/70 & 6/70 \\
    \texttt{zdtad} & 22/70 & 21/70 & 15/70 \\
    \texttt{joker} & 23/70 & 14/70 & 8/70 \\
    \texttt{firead} & 17/70 & 15/70 & 8/70 \\
    \texttt{blacklister} & 17/70 & 13/70 & 9/70 \\
    \texttt{gappusin} & 18/70 & 16/70 & 12/70 \\
    %\midrule 
    \cmidrule(){1-4}
    \textbf{Average} & 27\% & 22\% & 13\% \\ \bottomrule
\end{tabular}
}
\end{table}

\section{Conclusions and Future Work}
\label{sec:conclusions}
In this work, we evaluated existing problem‑space attacks for Android malware detection and showed that, despite claims, they remain largely ineffective. Coarse‑grained transplantation methods introduce excessive side effects and frequently fail to rebuild, while fine‑grained techniques either lack impact or break semantics. Moreover, prior work overestimates attack success, due to insufficient functionality testing that cannot properly assess the preservation of the manipulated \apk behavior.
To address these limitations, we introduced \practicaldroid, a framework for crafting adversarial APKs that are both practical (build-safe) and functional (semantics-preserving), providing the following contributions: (i) query‑efficient white‑ and black‑box attacks based on model‑influential transformations; (ii) fine‑grained, build‑safe manipulation of APIs, modules, Permissions, and URLs with minimal side effects; and (iii) a semantics‑preserving functionality test enforcing runtime equivalence via execution logs and API‑level traces. Our empirical results confirm the practical impact of \practicaldroid: (i) it achieves high evasion rates against multiple detectors in both white‑ and black‑box settings with only a few tens of queries and negligible side effects; (ii) the resulting adversarial APKs remain functional at runtime; and (iii) when evaluated against commercial scanners on \gls{VT}, it significantly reduces detections, demonstrating that practical and semantics‑preserving problem‑space attacks are feasible at scale.
\practicaldroid enables several promising directions for future work. First, thanks to its query efficiency, it can serve as a foundation for developing \textit{adaptive} attacks against stateful‑protection and rate‑limiting defenses~\cite{chen2020stateful,debenedetti2024evading}. Another natural extension is to broaden its manipulation set to target detectors that rely on \textit{dynamic} features, modifying execution‑level behaviors beyond those considered by existing problem‑space attacks.

\section{Ethical  Considerations}
\practicaldroid aims to improve the robustness evaluation of machine-learning Android malware detectors. Our goal is to highlight existing vulnerabilities and encourage the development of more robust and secure detection systems. For this reason, we have chosen to publicly release our attack implementation to foster reproducibility, fair benchmarking, and the design of improved countermeasures. This approach follows established practices in the security and machine learning communities, where openness and responsible disclosure are key to strengthening defenses (cf.~\cite{li2020adversarial, demetrio2021functionality}).
However, we acknowledge the dual-use nature of this work. Adversarial techniques designed for research can also be misused to evade real-world detection systems. To mitigate this risk, we decided to (i) document the intended purpose of the tool—enhancing robustness, not enabling malicious activity; (ii) include explicit disclaimers and responsible usage guidelines; and (iii) release the code under a restrictive license prohibiting unethical or harmful applications. These measures, combined with transparency and community oversight, aim to maximize the scientific and defensive value of our contribution while minimizing potential misuse.

\section*{Acknowledgment}
This work has been partly supported by the EU-funded Horizon Europe projects ELSA (GA no. 101070617) and CoEvolution (GA no. 101168560); and by the projects SERICS (PE00000014) and FAIR (PE00000013) under the MUR National Recovery and Resilience Plan funded by the European Union - NextGenerationEU. This work was carried out while C. Scano was enrolled in the Italian National Doctorate on AI run by the Sapienza University of Rome in collaboration with the University of Cagliari.

\bibliographystyle{IEEEtran}
\bibliography{bibliography}

\begin{appendices}

\section{Problem-space Attacks Details}

\begin{table*}[htbp]
\caption{Performance metrics for considered detectors, on both \apg and \elsa datasets.}
\centering
\resizebox{\textwidth}{!}{%
\begin{tabular}{@{\extracolsep{\fill}} l c c c c c c c c c c}
    \toprule
    \multirow{2}{*}[-0.2em]{\centering \textbf{Metric}} & \multicolumn{6}
    {c}{\textbf{APG DS}} & \multicolumn{4}{c}{\textbf{ELSA DS}} \\
    \cmidrule(lr){2-7} \cmidrule(lr){8-11}
    & \textbf{\drebin} & \textbf{\drebinred} & \textbf{\secsvm} & \textbf{\secsvmred} & \textbf{\apigraph} & \textbf{\mamadroid} & 
    \textbf{\drebin} & \textbf{\secsvm} & \textbf{\apigraph} & \textbf{\mamadroid} \\
    \midrule
    F1-score & 78.2\% & 79.63\% & 76.98\% & 75.85\% & 79.76\% & 78.34\% & 86.48\% & 85.09\% & 85.89\% & 51.86\% \\[0.2em]
    Precision & 72.5\% & 72.89\% & 72.03\% & 73.19\% & 72.81\% & 76.31\% & 97.72\% & 94\% & 96.79\% & 91.68\% \\[0.2em]
    Recall & 84.86\% & 87.74\% & 82.66\% & 78.71\% & 88.17\% & 80.49\% & 77.28\% & 75.36\% & 77.2\% & 36.16\% \\[0.2em]
    \bottomrule
\end{tabular}
}
\smallskip
\label{tab:detector_performance}
\end{table*}

\subsection{Requirements for Problem-space Attacks}
\label{appendix:problem-space}
Pierazzi et al.~\cite{pierazzi2020intriguing} formalized four key requirements for problem-space attacks:
\begin{enumerate}
    \item[(i)] \textit{practical manipulations}, \ie the APK changes must be feasible under Android’s structural and syntactic constraints, even if they may introduce unintended side‑effect features;
    \item[(ii)] \textit{semantics preservation}, \ie the modified APK continues to exhibit its intended (malicious) behavior;
    \item[(iii)] \textit{robustness to preprocessing}, \ie the manipulations cannot be easily removed or neutralized by trivial preprocessing mitigation measures; and
    \item[(iv)] \textit{plausibility}, \ie the modified APK remains inconspicuous to both users and analysts.
\end{enumerate}

\myparagraph{Robustness to Preprocessing} While we agree that problem‑space attacks should rely on practical manipulations and preserve app semantics, we argue that implementing robust preprocessing mechanisms may not be as trivial as often claimed and, thus, it should not be considered a strict requirement. While prior work claims that ``trivial'' preprocessing steps may neutralize certain types of manipulation, no machine-learning malware detector to date implements them~\cite{arp2014drebin,mariconti2017mamadroid}. As the complexity of identifying unreachable code, callbacks, or reflection instructions via static analysis makes their removal non-trivial and prone to error, in practice, such manipulations remain effective, making this requirement poorly specified and rarely supported. Nevertheless, while our manipulations are mainly designed to be practical (build-safe), they can be hardened with similar tricks from previous work to demonstrate their robustness to preprocessing.

\myparagraph{Plausibility} As for robustness to preprocessing, we argue that plausibility should not be considered a strict requirement, too. In the context of adversarial images, Gilmer et al.~\cite{gilmer18} observed that \textit{``no compelling example where imperceptibility is required''} could be found, showing that the goal of evading a classifier is fundamentally different from that of also fooling a human.\footnote{It is important to distinguish imperceptibility from semantics preservation. An attack can introduce visible perturbations to an image without altering its semantics, so a human observer can still correctly recognize its content.} Similarly, manipulated \apks may only need to bypass the detector, without necessarily appearing inconspicuous to malware analysts---which is indeed a different goal.
Furthermore, the notion of plausibility in prior work has never been empirically validated. Pierazzi et al.~\cite{pierazzi2020intriguing} claimed that techniques such as opaque predicates and software transplantation inherently satisfy this requirement, whereas obfuscation does not. Beyond the fact that opaque predicates themselves introduce obfuscation (e.g., hiding false if clauses to inject dead code), this claim has not been subjected to human studies or user evaluations to verify whether the resulting APKs would indeed appear inconspicuous to analysts. The plausibility requirement remains thus largely theoretical, and we decided not to enforce it as a strict requirement.

\subsection{Manipulation Types}\label{app:manipulation:types}
Prior work on Android malware typically categorizes APK manipulations by the program representation they affect, ranging from high‑level declarative metadata to low‑level code.

\myparagraph{Manifest‑level Manipulations}
These manipulations target the Android manifest, including injection or modification of App Modules (\ie Activities, Services, Receivers, Providers), Permissions, Hardware Features, and Intent Filters~\cite{li2020adversarial}, and directly affect high‑level static features but are tightly constrained by Android’s consistency and validation rules.

\myparagraph{DEX bytecode–level Manipulations}
They target Dalvik bytecode by injecting or modifying classes, methods, API calls, and embedded constants (e.g., URLs, IP addresses), with changes ranging from coarse‑grained (e.g., transplanting entire classes or modules)~\cite{pierazzi2020intriguing, yang2017malware} or fine‑grained (e.g., inserting individual API calls or rewriting specific instructions)~\cite{li2020adversarial}.

\myparagraph{Control‑Flow Graph (CFG) Manipulations}
CFG‑level manipulations alter execution structure without necessarily adding or removing semantic components. Typical examples include basic-block reordering, opaque predicates, function inlining or duplication, and control‑flow flattening. Prior work (e.g., HRAT~\cite{zhao2021structural}) leverages such transformations to perturb structural features or increase code complexity; however, these operations usually preserve all original components, providing only partial obfuscation or no removal of malicious capability.

\myparagraph{Native‑code (JNI) Manipulations}
They affect compiled native libraries (ELF binaries) accessed through Java Native Interface (JNI), and include injecting benign native libraries, modifying native call patterns, and renaming symbols~\cite{Ruggia25_PrivSec}. Even if native code is a relevant attack surface, such manipulations are less explored due to architectural dependencies, limited tooling, and a higher risk of breaking functionality.

\myparagraph{Resource‑level Manipulations}
They modify \apk assets such as layout files, images, and resource string tables without altering app semantics, but they may still influence detectors that rely on metadata, resource statistics, and strings.

Overall, existing attacks differ both in which representations they manipulate and in the granularity of the manipulation: coarse‑grained approaches simultaneously affect manifest, bytecode, FCG, resources, and sometimes native code, while fine‑grained approaches restrict themselves to a narrow subset to reduce side effects and preserve functionality.

\subsection{APK Component Examples}\label{app:app_components}

We report concrete examples of \apk components that are manipulated by problem-space attacks to evade Android malware detectors, organized by the categories in~\autoref{subsubsec:adv_apk:apk_components}.
% , and shown in the examples in \autoref{fig:manifest_components}-\ref{fig:smali_components}

\myparagraph{\comp{1}: Application Modules} They correspond to the class names of \apk entry points (Activities, Services, Receivers, and Providers), declared within the manifest with specific tags, \ie {\tt <activity>}, {\tt <service>}, {\tt <receiver>}, {\tt <provider>}, and implemented inside the DEX bytecode.

\myparagraph{\comp{2}: Hardware Features} They are declared within the manifest with the tag {\tt <uses-features>} and declare the permission to use specific hardware (e.g., microphone, camera).
%and its name starts with {\tt android.harware}.

\myparagraph{\comp{3}: Permissions} They are declared within the manifest with the tag {\tt <uses-permission>} and identify the app software capabilities (e.g., sending SMS, or accessing storage). 
%and its name starts with {\tt android.permission}.
   
\myparagraph{\comp{4}: Intent Filters} They are reported within the manifest and describe the external behaviors of app modules invoked at runtime. They are identified by the XML tag {\tt <intent-filter>}, and the {\tt <action>} name that identifies the behavior, declared within an Application Module.

\myparagraph{\comp{5}: API calls} They are used by the DEX bytecode to interact with the Android OS and implement \apk functionality. They are invoked in the smali code with instructions like {\tt invoke-*}. In our case, we consider only framework's APIs.

\myparagraph{\comp{6}: Strings} They include Strings (e.g., URLs, IPs) initialized within the DEX bytecode. They are declared with the instruction {\tt const-string} that pushes a constant string into a register. 
In our attack, we consider only IP addresses and URLs as strings to be modified.

\subsection{Repackaging}
\label{app:problem-space:repackaging}
Repackaging enables the extraction and manipulation of an app’s manifest, code, and resources. Most manipulation operators are applicable only to APKs that can be safely repackaged without introducing artifacts that invalidate the DEX code or are unrelated to the intended modifications; consequently, apps protected by packing or advanced obfuscation techniques are typically excluded from adversarial manipulation.
In practice, tools such as \texttt{Jadx}~\cite{Jadx} and \texttt{ApkTool}~\cite{ApkTool} are used to decompile DEX files, recovering the original Java code when possible or lifting it to smali~\cite{smali}, an assembly‑like language for the Android Virtual Machine. The resulting code and resources can then be modified and repackaged, producing patched versions of the original APKs that can be redistributed and installed, provided the applied changes do not compromise the app's integrity. Finally, the repackaged APKs must be \emph{re‑signed} using \texttt{apksigner}~\cite{Apksigner}, as Android requires each APK to include a valid digital certificate to verify its integrity and allow installation on a device.

\section{Reproducibility and Implementation Issues}

\subsection{Pierazzi et al. Implementation Issues}\label{app:manipulation:pierazzi}
We outline here the issues we found with the approach by Pierazzi~\etal~\cite{pierazzi2020intriguing} that hinder the \apks from being built or functioning correctly.

First, as also shown in previous work~\cite{olszewski23getin}, reproducibility issues mainly due to incomplete documentation and configuration details, prevent the attack from working out of the box. In particular, we observed that the attack's operational stability is significantly influenced by the interplay between the versions of the manipulated \apk SDK, FlowDroid, Android Build Tools, and Platform Tools.
After contacting the authors and trying several settings, we achieved optimal results using Build Tools 23.0.1 and Platform Tools 23.

Second, the original opaque predicate within which the benign code is injected was compiled into a control flow structure that Soot could not recognize as a valid injection point, preventing transplantation from working correctly. A minor modification to the conditional expression resolved the issue without affecting the predicate's runtime behavior.

Third, after selecting all organs to inject, FlowDroid identifies their dependencies, which are subsequently added to the app as stub classes, even when a filtering step is supposed to remove unnecessary components. However, this is not effective, and the approach also fails to discard malformed dependency classes with invalid access flags, \eg 0x20601, which incorrectly assigns the {\tt synchronized} modifier to those classes. 
To verify this, we recompiled the \apks without the faulty classes and, as specified in~\autoref{subsubsec:evaluation:functionality-check-wb}, launched them without fatal errors from the Android Verifier.

Overall, these issues highlight the approach's lack of robustness, which fails to adequately handle edge cases or ensure the correctness of injected components, ultimately leading to unstable, unreliable \apk builds.

\begin{figure}[t]
    \centering
\begin{subfigure}{0.46\textwidth}
\begin{lstlisting}[style=xml,escapechar=!]
<?xml version="1.0" encoding="utf-8"?>
<manifest package="com.nd.android.pandah" [...]>
    [...]
    <application android:label="@string/application_name" android:name="com.nd.hilaunch.LauncherApp2">
    [...]
    <activity android:name="com.nd.hilaunc.Launcher"/>
    <acvitity android:name="!\codehl{blue!20}{R345678shgd}!" 
        android:label=""> !\label{line:act-inj}!
    <activity android:name="!\codehl{blue!20}{A2b6bdaa4}!"> !\label{line:faulty-act-inj-ade}!
        <intent-filter>
        <action android:name=
            "android.intent.action.VIEW"/> !\label{line:intent}! 
        </intent-filter>
    </activity>!\label{line:end-act-inj-ade}!
    [...]
    </application>
</manifest>
\end{lstlisting}\end{subfigure}
\vspace{-2\baselineskip}
\caption{\ade App-Module Injection. Injected App Modules components are highlighted in blue ($C_1$).}\label{fig:mod_inj-ade}
\end{figure}

\subsection{\ade Manipulation Issues}\label{app:manipulation:ade}
We outline here the \ade implementations that are brittle, causing the perturbed apps to crash due to Android Verifier errors or runtime exceptions.

\myparagraph{\underline{Component Obfuscation}}
\noindent\textbf{App-Module Class Renaming}. The approach encodes the class identifier and tokens into the package name and propagates this transformation to all matching string occurrences in both manifest and smali.
As a result, class names are only partially modified in the smali code, and the renamed identifier appears in the method body. However, the corresponding \texttt{.class} definitions are left unchanged, yielding to {\tt ClassNotFoundException} as the referenced classes are not actually defined.

\noindent\textbf{API Indirection and Reflection}. \ade implements API Reflection only, without considering other techniques like Indirection. To do so, it enforces the usage of {\tt getMethod} whichever the API to be reflected is. While being syntactically correct, it raises problems when the attack chooses inaccessible methods, \ie private, which cannot be resolved using {\tt getMethod} as it can be used only with public APIs. As a result, the reflective lookup fails at runtime, raising a {\tt NoSuch\-MethodException}, invalidating functionality.

\myparagraph{\underline{Component Injection}}
\noindent\textbf{App-Module Injection}. \autoref{fig:mod_inj-ade} outlines two examples of Activity injections by \ade that suffer from implementation issues. First, when injecting manifest components \ade does not detect that it is faulty and inserts it into the manifest, leading to compilation errors (\eg in line~\ref{line:act-inj} {\tt label} attribute is empty, but it should contain a valid value).
Second, since the Activities are injected only at the manifest level, there is no corresponding code within the DEX. As a result, when such an Activity is launched (\eg when an app issues the corresponding Intent), the app may fail since the Activity class is missing, raising errors.

\noindent\textbf{String Injection}. The major problem with this kind of manipulation is that \ade employs inaccessible registers, \eg {\tt p0} that holds the Java {\tt this} and cannot store string objects, or registers that are not previosly declared in the method register's set. As a result, the string injection cause the Android Verifier to reject the app with a fatal error invalidating its functionality.

% \begin{figure}[t]
%     \centering
%     \begin{subfigure}{0.46\textwidth}
%     \begin{lstlisting}[style=smali,escapechar=?]
% .method public constructor <init>(Lcom/zxml/image/pc8;)V
%     .registers 2 ?\label{line:reg-declare}?
%     const-string p0, "?\textcolor{blue}{http://192.168.1.100}?" ?\label{line:str-p0}?
%     const-string v3, "?\textcolor{blue}{https://gjapplog.ucweb.com}?" ?\label{line:str-v}?
%     iput-object p1,p0,Ldj;->a:Lcom/zxml/image/pc8;
%     invoke-direct {p0}, Ljava/lang/Object;-><init>()V
%     return-void
% .end method
% \end{lstlisting}\end{subfigure}
%     \vspace{-2\baselineskip}
%     \caption{\ade URL (String) injection. Injected URLs ($C_6$) are highlighted in blue.}
%     \label{fig:str-injection-ade}
% \end{figure}

\subsection{\practicaldroid Manipulation Details}\label{app:manipulation:our}
We outline here the implementation details of the manipulations presented in~\autoref{subsec:methodology:func_pres_manipulation}, explaining how they preserve functionality and differ from prior work.

\myparagraph{\underline{Component Obfuscation}}
\noindent\textbf{App-Module Class Renaming}.
Unlike \ade, to avoid inconsistent and partial renaming, we encode only the class identifiers within the manifest (\autoref{fig:class_rename-our:manifest}), while preserving the app package tokens.
After encoding the Activity class name in the manifest (\texttt{com.pc3ef4.p75645.\-p72b28} in line~\ref{line:encoding}), we consistently rename every occurrence of the same class within the smali code, without partial renaming of other classes even though it contains similar tokens (\ie~{\tt com}). 
%\textcolor{red}{anche qui dovete essere un po' più precisi indicando qual è il package e qual è il nome della classe.}.

\begin{figure}[t]
    \centering
    \begin{subfigure}{0.46\textwidth}
    \begin{lstlisting}[style=xml, escapechar=!]
<?xml version="1.0" encoding="utf-8"?>
<manifest package="com.android.pandahome2" [...]>
    [...]
    <application android:label="@string/application_name" android:name="com.hilauncherdev.launcher.LauncherTinkerApplication">
    [...]
    <activity android:name="!\codehl{blue!20}{com.pc3ef4.p75645.p72b28}!"/> !\label{line:encoding}!
    [...]
    </application>
</manifest>
\end{lstlisting}
\end{subfigure}
        \vspace{-2\baselineskip}
        \caption{App Module ($C_1$) renaming in the manifest (line~\ref{line:encoding}) without encoding the tokens that are also within the package name.}
        \label{fig:class_rename-our:manifest}
\end{figure}

\begin{figure}[h]
    \centering
    \begin{subfigure}{0.46\textwidth}
        \centering
        \begin{lstlisting}[style=smali, escapechar=!]
.method public a(I)Landroid/graphics/drawable/Drawable;
    .registers 6
    [...]
    move-result v0
    invoke-virtual {v1, v0}, !\codehl{blue!20}{Landroid/content/res/Resources;->getDrawable(I)}!
        !\codehl{blue!20}{Landroid/graphics/drawable/Drawable;}!
    move-result v0
    [...]
.end method\end{lstlisting}
        \vspace{-2\baselineskip}
        \caption{Plain version of the {\tt a} function.}\label{fig:call_ind:plain}
    \end{subfigure}
    \hfill
    \begin{subfigure}{0.46\textwidth}
        \begin{lstlisting}[style=smali,escapechar=?]
.method public static uQtOtnIMMiVSwoJq(Landroid/content/res/Resources;I)Landroid/graphics/drawable/Drawable;
    .locals 1
    invoke-virtual {p0, p1}, ?\codehl{blue!20}{Landroid/content/res/Resources;->getDrawable(I)}?
        ?\codehl{blue!20}{Landroid/graphics/drawable/Drawable;}? ?\label{line:call_ind:invoke}?
    move-result v0
    return v0
.end method

.method public a(I)Landroid/graphics/drawable/Drawable;
    .registers 6
    [...]
    move-result v0
    invoke-static {v1, v0}, Lcom/google/android/gms/plus/e;->uQtOtnIMMiVSwoJq(Landroid/content/res/Resources;I)Landroid/graphics/drawable/Drawable;
    move-result v0
    [...]
.end method\end{lstlisting} 
        \vspace{-2\baselineskip}
        \caption{Obfuscated version of the {\tt a} function through Call Indirection.}\label{fig:call_ind:obf}
    \end{subfigure}
    \caption{Obfuscation through Call Indirection. (a) shows the smali code of the original function {\tt a} with the call to the target API in blue ($C_5$), while (b) shows the obfuscated counterpart.}
    \label{fig:call_ind}
\end{figure}

\noindent\textbf{API Indirection}. Unlike \ade, we also propose API indirection to obfuscate APIs. As shown in~\autoref{fig:call_ind}, we obfuscate the original call to \texttt{getDrawable}  (Fig.~\ref{fig:call_ind:plain}), with wrapper method (Fig.~\ref{fig:call_ind:obf}) invoking the API (line~\ref{line:call_ind:invoke}), altering the \gls{FCG} with an additional node. To maintain the app's functionality, the new method accepts the same parameters as the obfuscated one and the corresponding object to correctly invoke it.

\noindent\textbf{API Reflection}.~\autoref{fig:reflection} shows an example of API Reflection. We leverage an additional class \texttt{ApiReflection} (\autoref{fig:reflection:obf}), and its {\tt obfuscate} method, which takes in input an integer (\ie the identifier of the call in the array), and the parameters passed to the target method to invoke (line~\ref{line:invoke}). As in the case of \ade, we employ {\tt getMethod} (line~\ref{line:getMethod}) from {\tt Ljava/lang/reflect/Method} class, but differently from it, we employ only public methods from public classes, not to raise errors of inaccessibility.

\begin{figure}[t]
    \centering
    \begin{subfigure}{0.46\textwidth}
        \centering
        \begin{lstlisting}[style=smali, escapechar=?]
.method public b(Landroid/os/Parcel;)Lcom/google/android/gms/wallet/NotifyTransactionStatusRequest;
    .registers 9
    [...]
    invoke-virtual {v1, v2}, ?\codehl{blue!20}{Landroid/webkit/WebSettings;->setSupportZoom(Z)V}?
    [...]
.end method\end{lstlisting}
        \vspace{-2\baselineskip}
        \caption{Plain version of the {\tt b} function.}\label{fig:reflection:plain}
    \end{subfigure}
    \hfill
    \begin{subfigure}{0.46\textwidth}
        \begin{lstlisting}[style=smali,escapechar=?]
.method public b(Landroid/os/Parcel;)Lcom/google/android/gms/wallet/NotifyTransactionStatusRequest;
    .registers 9
    [...]
    invoke-static {v8, v0, v6}, Lcom/reflect/ApiReflection;->obfuscate(ILjava/lang/Object;[Ljava/lang/Object;)Ljava/lang/Object;
.end method

.class public Lcom/reflect/ApiReflection;
.super Ljava/lang/Object;

.method static constructor <clinit>()V
    [...]
    const-class v2, ?\codehl{blue!20}{Landroid/webkit/WebSettings;}?
    const-string v3, "?\codehl{blue!20}{setSupportZoom}?"
    invoke-virtual {v2, v3, v1}, Ljava/lang/Class;->getMethod(Ljava/lang/String;[Ljava/lang/Class;)Ljava/lang/reflect/Method;?\label{line:getMethod}?
    move-result-object v1
    sget-object v2, Lcom/reflect/ApiReflection;->obfuscatedMethods:Ljava/util/List;
    invoke-interface {v2, v1}, Ljava/util/List;->add(Ljava/lang/Object;)Z
.end method

.method public static obfuscate(ILjava/lang/Object;[Ljava/lang/Object;)Ljava/lang/Object;
    sget-object v0, Lcom/reflect/ApiReflection;->obfuscatedMethods:Ljava/util/List;
    invoke-interface {v0, p0}, Ljava/util/List;->get(I)Ljava/lang/Object;
    move-result-object p0
    check-cast p0, Ljava/lang/reflect/Method;
    invoke-virtual {p0, p1, p2}, ?\codehl{red!20}{Ljava/lang/reflect/Method;->}?
    ?\codehl{red!20}{invoke(Ljava/lang/Object;[Ljava/lang/Object;)}?
    ?\codehl{red!20}{Ljava/lang/Object;}? ?\label{line:invoke}?
    [...]
.end method
\end{lstlisting} 
        \vspace{-2\baselineskip}
        \caption{Obfuscated version of the {\tt b} function through Reflection.}\label{fig:reflection:obf}
    \end{subfigure}
    \caption{\practicaldroid API Reflection. (a) shows the smali code of the original function {\tt b} with the call to the target API in blue ($C_5$), while (b) shows the obfuscated counterpart with the corresponding side-effect in red.}
    \label{fig:reflection}
\end{figure}

\noindent\textbf{String Encryption}. \practicaldroid encrypts strings to hide the URL inside the method body. In particular, we leverage a {\tt DecryptString} classes that have all necessary methods to descript the method name to dynamically reconstruct the original value at runtime.  This transformation preserves the original semantics, while preventing static analyzers from directly observing sensitive string literals.

% \begin{figure}[ht]
%     \centering
%     \begin{subfigure}{0.46\textwidth}
%         \centering
%         \begin{lstlisting}[style=smali,escapechar=?]
% .method public b()Ljava/lang/String;
%     .registers 2
%     const-string v0, "?\textcolor{blue}{http://127.0.0.1}?" ?\label{line:string-p}?
%     return-object v0
% .end method
% \end{lstlisting}
%         \vspace{-2\baselineskip}
%         \caption{Plain version of the URL declared within the {\tt b} method.}\label{fig:encrypt:plain}
%     \end{subfigure}
%     \hfill
%     \begin{subfigure}{0.46\textwidth}
%         \begin{lstlisting}[style=smali,escapechar=?]
% .method public b()Ljava/lang/String;
%     const-string/jumbo v0, "58b1675abc338ec4664bb121d6d393c1d57e98918d803
%         a260cfbaebfed38c4b6" ?\label{line:string-enc}?
%     invoke-static {v0}, Lcom/decryptstringmanager/DecryptString;->decryptString(Ljava/lang/String;)Ljava/lang/String;
%     move-result-object v0
%     return-object v0
% .end method
% \end{lstlisting} 
%         \vspace{-2\baselineskip}
%         \caption{Encrypted version of the URL declared within the {\tt b} method using runtime {\tt decryptString}.}\label{fig:encrypt:obf}
%     \end{subfigure}
%     \caption{\practicaldroid URL (String) Encryption. (a) shows the original version of the String component in blue ($C_6$),  while (b) shows the encrypted counterpart of the URL.}
%     \label{fig:encrypt}
% \end{figure}
% Lines 4 and 5 are the injected
% lines, with the blue tokens highlighting the injected String
% components (��6).

\myparagraph{\underline{Component Injection}}
\noindent\textbf{App-Module Injection}.~\autoref{fig:mod_inj-our} depicts an example of Activity injection by \practicaldroid. To overcome the implementation limitations of \ade, we first inject only the Activity name without adding any additional attribute from the benign manifest that may raise build errors (line~\ref{line:act-inj-simple}), then we set the {\tt enable} attribute to false so that the Android system treats the Activity as disabled at runtime; therefore, the Intents are not resolved, and the injected Activity, \ie without corresponding code, cannot be launched.

\begin{figure}[t]
    \centering
\begin{subfigure}{0.46\textwidth}
\begin{lstlisting}[style=xml,escapechar=!]
<?xml version="1.0" encoding="utf-8"?>
<manifest package="com.nd.android.pandah" [...]>
[...]
<uses-permission android:name="!\codehl{blue!20}{android.permission.}!
    !\codehl{blue!20}{SEND\_SMS}!"/> !\label{line:perm}!
<uses-feature android:name=
    "!\codehl{blue!20}{android.hardware.audio}!"/> !\label{line:hw}!
<application android:label="@string/application_name" android:name="com.nd.hilauncherdev.Launch">
[...]
<acvitity android:name="!\codehl{blue!20}{R345678shgd}!"
    android:enable="false" !\label{line:act-inj-simple}!
<activity android:name="!\codehl{blue!20}{A2b6bdaa4}!"
    android:enable="false"> !\label{line:act-inj-our}!
    <intent-filter>
    <action android:name="!\codehl{blue!20}{android.intent.action.SEND}!"/> !\label{line:intent}! 
    </intent-filter>
</activity>!\label{line:finish-act-inj-our}!
[...]
</application>
</manifest>
\end{lstlisting}\end{subfigure}
\vspace{-2\baselineskip}
\caption{\practicaldroid manifest components Injection, \ie Permissions ($C_3$), Hardware Features ($C_2$), App Modules ($C_1$), and Intent Filers ($C_4$).
}\label{fig:mod_inj-our}
\end{figure}

\begin{figure}[t]
    \centering
    \begin{subfigure}{0.46\textwidth}
    \begin{lstlisting}[style=smali,escapechar=?]
.class public Lcom/apiinjectionmanager/ApiInjection;
.super Ljava/lang/Object;

.method public static inject()V
    .registers 6
    const/4 v0, 0x1
    const/4 v1, 0x0
    /* Logging */
    if-nez v0, :impossible
    invoke-static {v2},?\codehl{blue!20}{Landroid/os/Binder;->getCallingPid()I}? ?\label{line:api_inj}?
    :impossible
    return-void
.end method\end{lstlisting}
\end{subfigure}
        \vspace{-2\baselineskip}
        \caption{\practicaldroid API Injection. \texttt{inject} holds the injected API (\comp{5}).}
        \label{fig:api-injection}             
\end{figure}

\noindent\textbf{Hardware Features} and \textbf{Permission Injection}.~\autoref{fig:mod_inj-our} shows an example of injection of hardware feature (line~\ref{line:hw}) and permission (line~\ref{line:perm}).
In particular, our implementation takes into account only components from the Android framework starting with {\tt android.permission} and {\tt android.hardware}, so that apps does not need the corresponding declaration.

% \begin{figure}[ht]
%     \centering
%     \begin{subfigure}{0.46\textwidth}
%     \begin{lstlisting}[style=xml,escapechar=!]
% <?xml version="1.0" encoding="utf-8"?>
% <manifest package="com.nd.android.pandahome2" [...]>
% <uses-permission android:name="!\textcolor{blue}{android.permission.}!
%     !\textcolor{blue}{SEND\_SMS}!"/> !\label{line:perm}!
% <uses-permission android:name="android.permission.ACCESS_FINE_LOCATION"/>
% <uses-feature android:name=
%     "!\textcolor{blue}{android.hardware.microphone}!"/> !\label{line:hw}!
% <uses-feature android:name="android.hardware.bluetooth"/>
% <application android:label="@string/application_name" android:name="com.nd.hilauncherdev.launcher.LauncherTinkerApplication">
% [...]
% </application>
% </manifest>
% \end{lstlisting}\end{subfigure}
%     \vspace{-2\baselineskip}
%     \caption{\practicaldroid Hardware and Permission Injection. Injected Permissions ($C_3$) and Hardware Features ($C_2$) are highlighted in blue.}
%     \label{fig:manifest-inj}
% \end{figure}

\noindent\textbf{API Injection}. \autoref{fig:api-injection} depicts an example of API injection. New APIs are injected by introducing a new class whose {\tt void} {\tt inject} method contains the invocations of additional APIs within dead code. 
To preserve the functionality, (i)~we inject not only the target API (line~\ref{line:api_inj}) and (ii)~we employ a {\tt void} method not to interfere with the app context when called.

\noindent\textbf{String Injection}. Rather than injecting string constants into an existing app method, violating register allocation constraints as done by \ade, we introduce a dedicated {\tt void} method that (i) safely allocates fresh registers and stores the injected strings locally, and (ii) does not return any value, so as not to interfere with the original app context when called. 
As a result, our implementation preserves the bytecode correctness and avoids Android Verifier errors.

\end{appendices}

\end{document}